\documentclass[12pt,english]{article}
\usepackage[T1]{fontenc}
\usepackage[latin9]{inputenc}
\usepackage{amsmath}
\usepackage{amssymb}

\makeatletter
\newcommand{\be}{\begin{equation}}\newcommand{\ee}{\end{equation}}\newcommand{\bea}{\begin{eqnarray}}\newcommand{\eea}{\end{eqnarray}}

\def\({\left(}\def\){\right)}\usepackage{epsfig}

\makeatother

\usepackage{babel}

\begin{document}

\title{Approximate solution to the CGHS field equations for two-dimensional evaporating black holes}

\author{Amos Ori \\ Department of Physics, Technion-Israel Institute of Technology, \\  Haifa 32000,  Israel }

 \maketitle
 
 \begin{abstract} 

Callan, Giddings, Harvey and Strominger (CGHS) previously introduced a two-dimensional semiclassical model of gravity coupled to a dilaton and to matter fields. Their model yields a system of field equations which may describe the formation of a black hole in gravitational collapse as well as its subsequent evaporation. Here we present an approximate analytical solution to the semiclassical CGHS field equations. This solution is constructed using the recently-introduced formalism of flux-conserving hyperbolic systems. We also explore the asymptotic behavior at the horizon of the evaporating black hole.

\end{abstract}


\section{Introduction}

The semiclassical theory of gravity treats spacetime geometry at the
classical level but allows quantum treatment of the various fields
which reside in spacetime. This theory asserts that a quantum field
living on a black-hole (BH) background will usually be endowed with
non-trivial fluxes of energy-momentum. These fluxes, represented by
the \emph{renormalized stress-Energy tensor} $\hat{T}_{\alpha\beta}$,
originate from the field's quantum fluctuations, and typically they
do not vanish even in the (incoming) vacuum state. The Hawking radiation
\cite{Hawking}, and the consequent black-hole (BH) evaporation,
are perhaps the most dramatic manifestations of these quantum fluxes. 

In the framework of semiclassical gravity the spacetime reacts to
the quantum fluxes via the Einstein equations, which now receive the
extra quantum contribution $\hat{T}_{\alpha\beta}$ at their right-hand
side. The mutual interaction between geometry and quantum fields thus
takes its usual General-Relativistic schematic form: The renormalized
stress-Energy tensor $\hat{T}_{\alpha\beta}$ (say in the vacuum state)
is dictated by the background geometry, and the latter is affected
by $\hat{T}_{\alpha\beta}$ through the Einstein equations. In principle,
this evolution scheme allows systematic investigation of the spacetime
of an evaporating BH.

It turns out, however, that the calculation of $\hat{T}_{\alpha\beta}(x)$
for a prescribed background metric $g_{\alpha\beta}(x)$ is an extremely
hard task in four dimensions. Nevertheless, in two-dimensional (2D)
gravity the situation is remarkably simpler. There are two energy-momentum
conservation equations, and the trace $\hat{T}_{\alpha}^{\alpha}$
is also known (the "trace-anomaly"). These three pieces of information
are just enough for determining the three unknown components of $\hat{T}_{\alpha\beta}$.
It is therefore possible to implement the 
evolution scheme outlined above
in 2D-gravity, and to formulate a closed system of field equations
which describe the combined evolution of both spacetime and quantum fields.

Callan, Giddings, Harvey and Strominger (CGHS) \cite{CGHS} introduced
a formalism of 2D gravity in which the metric is coupled to a dilaton
field $\phi$ and to a large number $N$ of identical massless scalar
fields. They added to the classical action an effective term $\propto N$
which gives rise to the semiclassical trace-anomaly contribution,
and thereby automatically incorporates the renormalized stress-Energy
tensor $\hat{T}_{\alpha\beta}$ into spacetime dynamics. The evolution
of spacetime and fields is then described by a closed system of field
equations. They considered the scenario in which a BH forms in the
gravitational collapse of a thin massive shell, and then evaporates
by emitting Hawking radiation. Their main goal was to reveal the end-state
of the evaporation process, in order to address the information puzzle. 

The general analytical solution to the CGHS field equations is not
known. Nevertheless, these equations can be explored analytically \cite{RST} as well as numerically \cite{Piran,Dori,Pretorius}. 
The global structure which emerges from these studies is depicted in Fig. 1: 
The shell collapse leads to the formation of a BH. 
A spacelike singularity forms inside the BH, at a certain
critical value of the dilaton \cite{RST}. (The BH interior may be identified
as the set of all events from which all future-directed causal curves
hit the singularity.) The BH interior and exterior are separated by an outgoing null ray, which serves as an event horizon. Also an apparent horizon forms along a timelike line outside the BH (it is characterized by a local minimum of $R\equiv e^{-2\phi}$ along outgoing null rays).  The event and apparent horizons are denoted by "EH" and "AH" in Fig. 1.
Both horizons steadily "shrink" it time (namely $R$ decreases), exhibiting the BH evaporation process. At a certain point
the apparent horizon intersects with the spacelike singularity (and with the event horizon). This intersection  event (denoted "P" in Fig. 1) appears to be a naked singularity, visible to far asymptotic
observers. It may be regarded as the "end of evaporation" point.


\begin{figure}[h]
\begin{center}
\includegraphics[scale=0.6]{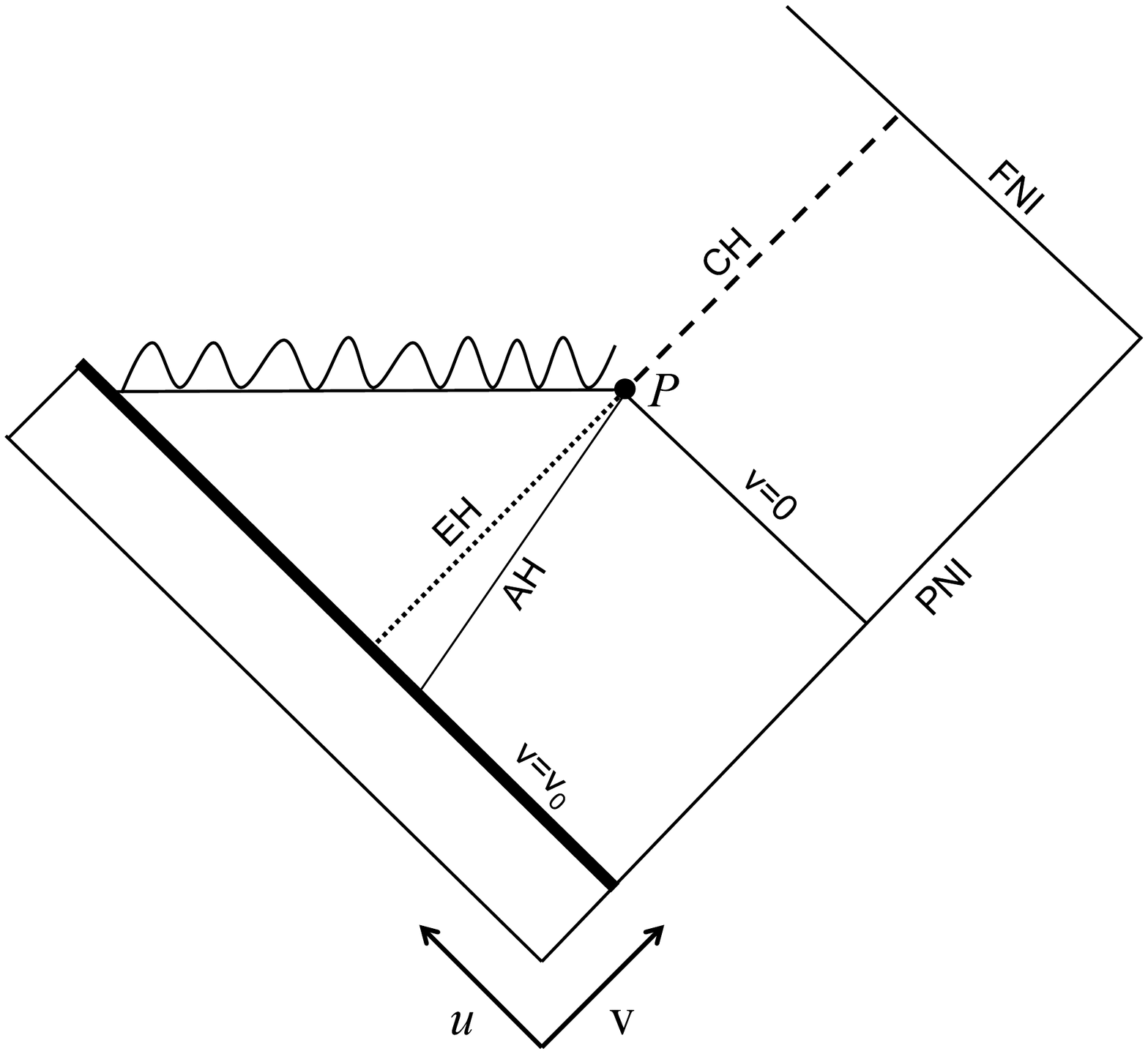}
\caption{ Penrose diagram describing the formation and evaporation of a two-dimensional CGHS black hole. The black hole forms by the collapse of a thin shell of macroscopic mass $M_{0}$. The collapsing shell is located at $v=v_{0}\equiv-M_{0}/q$. The lines denoted "EH", "AH", and "CH" respectively represent the event horizon, apparent horizon, and Cauchy horizon. }
\label{fig1}
\end{center}
\end{figure}


The spacelike singularity which develops inside a CGHS evaporating
BH was first noticed by Russo, Susskind, and Thorlacius \cite{RST}.
Its local structure was recently studied in some detail \cite{DO},
by employing the homogeneous approximation. This singularity marks
the boundary of predictability of the semiclassical CGHS formalism.
About two years ago Ashtekar, Taveras and Varadarajan \cite{Ashtekar}
proposed a quantized version of the CGHS model, in which the dilaton
and metric are elevated to quantum operators. This quantum formulation
of the problem seems to resolve the semiclassical singularity inside
the BH \cite{Ashtekar}, and thereby to shed new light on the information puzzle.
In a very recent paper \cite{DO-1} we applied a more simplified,
"Minisuperspace-like", quantum approach to the spacetime singularity
inside a CGHS evaporating BH. We obtained a (bounce-like) extension
of semiclassical spacetime beyond the singularity.

The goal of this paper is to construct an approximate analytical solution
to the semiclassical CGHS field equations, which will satisfactorily
describe the BH formation and evaporation. This approximate solution
applies as long as the BH is macroscopic, and as long as the spacetime
region in consideration is "macroscopic" too---namely, $R$
is sufficiently large (i.e. not too close to the singularity). In
such macroscopic regions, the semiclassical effects are weak in a
local sense. Our solution is "first-order accurate" in this
local sense; Namely, the local errors in the second-order field-equation
operators (applied to the approximate solution) are quadratic in the
magnitude of local semiclassical effects (this is further discussed
in Sec. \ref{sec:Verification}). 

The method we use here for constructing our approximate solution is
based on the formalism of \emph{flux-conserving systems} \cite{Ori-flux}.
This formalism deals with a special class of semi-linear second-order
hyperbolic systems in two dimensions. This class was first introduced
in Ref. \cite{Ori-charged}, and was subsequently explored more systematically
in Ref. \cite{gorbonos} (see also \cite{second}). In particular,
a flux-conserving hyperbolic system admits a rich family of single-flux
solutions, to which we refer as \emph{Vaidya-like solutions} (see
Sec. \ref{sec:Redefinition}), and for which the field equations reduce
to an \emph{ordinary} differential equation (ODE). Very recently we
demonstrated \cite{Ori-flux} that after transforming to new field
variables, the semiclassical CGHS equations take a form which is approximately
flux-conserving. Here we take advantage of this property and use the
formalism of flux-conserving systems to construct the approximate analytic
solution to the CGHS field equations. 

Our approximate solution is presented in Sec. \ref{sec:final-solution}.
It involves a single function denoted by $H$, which is defined through
a certain ODE. In order to make practical use of this approximate
solution, one must have at his disposal the solution for this ODE.
In Appendix \ref{sec:Appendix-H} we provide an approximate analytical solution to this ODE. 

In Sec. \ref{sec:CGHS} we briefly summarize the CGHS model, expressing
the semiclassical field equations in convenient variables $R,S$ (already
used in Refs. \cite{Ori-charged,DO,Ori-flux}). In particular we discuss
the classical model of a collapsing shell---which in fact provides
the initial conditions to the semiclassical problem of BH evaporation.
In Sec. \ref{sec:Redefinition} we introduce the new field variables
($W,Z$), which turn the field equations into a more standard form
(with no first-order derivatives). Then we observe the approximate
flux-conserving system obtained in these new variables at the large-$R$
limit. In Sec. \ref{sec:Construction} we construct the ingoing Vaidya-like
solution (a "single-flux" ingoing solution) of the flux-conserving
system, which constitutes the core to our approximate solution. This
core solution is complemented in Sec. \ref{sec:perturb} by adding
to it a weak outgoing component. The extra outgoing component
must be added in order to correctly satisfy the initial conditions
at the collapsing shell (and it is this component which eventually
gives rise to the outgoing Hawking radiation). In Sec. \ref{sec:final-solution}
we summarize our approximate solution, and also present an alternative
approximate expression for $S$. In Sec. \ref{sec:Gauges} we introduce
two useful gauges: the "shifted-Kruskal" gauge, and the semiclassical
Eddington gauge. The validity of the constructed solution is verified
in Sec. \ref{sec:Verification}. We first describe the magnitude of
the local error in the field equations, which turns out to be quadratic
in the local magnitude of semiclassical effects (assumed to be a small
quantity). Subsequently we verify that the initial data at the collapsing
shell are precisely matched by our approximate solution, and the same
for the initial data at past-null-infinity (PNI).

In Sec. \ref{sec:Horizon} we discuss the horizon of the semiclassical
BH and a few related issues. The horizon (or "event-horizon")
is the outgoing null line which separates the BH interior and exterior.
We present the behavior of $R$ and $S$ along the event horizon and
also compute the influx $T_{vv}$ into the BH. We also discuss the
location of the apparent horizon (defined by a minimum of $R$ 
along an outgoing null ray as
mentioned above). Then in Sec. \ref{sec:PFNI} we analyze the asymptotic
behavior at PNI, and also remark briefly on the asymptotic behavior
at future null infinity (FNI). Finally, in Sec. \ref{sec:Summary}
we summarize our main results.

\section{Background: The CGHS model \label{sec:CGHS}}

\subsection{Action and field equations}

The CGHS model \cite{CGHS} consists of 2D gravity
coupled to a dilaton $\phi$ and to a large number $N\gg1$ of identical
(free, minimally-coupled, massless) scalar fields $f_{i}$. Throughout
this paper we express the metric in double-null coordinates $u,v$
for convenience, namely \begin{equation}
ds^{2}=-e^{2\rho}dudv.\label{eq:metric}\end{equation}
 The action then takes the form \begin{equation}
\frac{{\rm 1}}{\pi}\int d^{2}\sigma\left[e^{-2\phi}\left(-2\rho_{,uv}+4\phi_{,u}\phi_{,v}-\lambda^{2}e^{2\rho}\right)-\frac{1}{2}\sum\limits _{i=1}^{N}f_{i,u}f_{i,v}+K\rho_{,u}\rho_{,v}\right],\label{eq:Action}\end{equation}
where $K\equiv N/12$. The term $\lambda^{2}$ denotes a cosmological
constant. We shall set $\lambda=1$ throughout. This is achieved by
a change of variable $\rho \to \rho'=\rho+\ln(\lambda)$, which
annihilates $\lambda$ but does not affect the field equations otherwise.
This setting actually amounts to a choice of basic length unit. 

The scalar fields all satisfy the trivial field equation \begin{equation}
f_{i,uv}=0.\label{eq:fuv}\end{equation}
The CGHS scenario consists of an imploding thin massive shell of mass
$M_{0}$ which moves along an ingoing null line $v=const\equiv v_{0}$,
forming a black hole. This shell is composed of ($v$-derivatives
of) the scalar fields $f_{i}$, which are concentrated as a $\delta$-function-like
distribution at $v=v_{0}$. The solution at $v<v_{0}$ is the trivial,
flat, vacuum solution (see below). The main objective of this paper
is the semiclassical solution which takes place at $v>v_{0}$. By
assumption, in this region too no incoming scalar waves are present.
Therefore the solution of Eq. (\ref{eq:fuv}) throughout the relevant
domain $v>v_{0}$ (as well as at $v<v_{0}$) is \begin{equation}
f_{i}(u,v)=0.\label{eq:fSOLUTION}\end{equation}

The remaining field equations consist of two evolution equations for
the fields $\phi,\rho$, as well as two constraint equations (see
\cite{CGHS}). To bring these equations to a simpler form we define
new field variables\begin{equation}
R\equiv e^{-2\phi}\;,\;\;\; S\equiv2(\rho-\phi),\label{eq:RSdefinition}\end{equation}
following Ref. \cite{Ori-charged}. The field equations for $R$ and $S$
then take the form \begin{equation}
R_{,uv}=-e^{S}-K\rho_{,uv},\label{eq:Requation}\end{equation}
 \begin{equation}
S_{,uv}=K\rho_{,uv}/R,\label{eq:Sequation}\end{equation}
where \begin{equation}
\rho=(S-\ln R)/2\label{eq:rho}\end{equation}
is to be substituted. The constraint equations become (after substituting
$f_{i}=0$ for the scalar-fields) \begin{equation}
R_{,uu}-R_{,u}S_{,u}+\hat{T}_{uu}=0,\label{eq:Ruu}\end{equation}
\begin{equation}
R_{,vv}-R_{,v}S_{,v}+\hat{T}_{vv}=0,\label{eq:Rvv}\end{equation}
where $\hat{T}_{uu},\hat{T}_{vv}$ are the semiclassical fluxes in the
two null directions, given by \begin{equation}
\hat{T}_{uu}=K\left[\rho_{,uu}-\rho_{,u}^{2}+z_{u}(u)\right],\label{eq:Tuu}\end{equation}
\begin{equation}
\hat{T}_{vv}=K\left[\rho_{,vv}-\rho_{,v}^{2}+z_{v}(v)\right].\label{eq:Tvv}\end{equation}
The functions $z_{u}(u),z_{v}(v)$ are initial functions which encode
the information about the system's quantum state. Following CGHS we
consider here an incoming vacuum state at PNI (apart from the imploding
massive shell, which is presumably encoded in the fields $f_{i}$
already at the classical level). Correspondingly the functions $z_{u}(u),z_{v}(v)$
are determined by the requirement that $\hat{T}_{uu}$ and $\hat{T}_{vv}$
vanish at PNI. 
\footnote{In a 2D spacetime there are two sectors of PNI, a "right PNI" and
a "left PNI". $z_{u}(u)$ is defined by the
demand $\hat{T}_{uu}=0$ at left PNI, and $z_{v}(v)$ by demanding
$\hat{T}_{vv}=0$ at right PNI. Similarly there are two sectors of
FNI, a right one and a left one. Throughout this
paper, by "PNI" and "FNI" we shall always refer to the right
sectors of PNI and FNI (unless stated otherwise), as is also illustrated in Fig. 1. }

It will sometimes be useful to re-express the system of evolution
equations (\ref{eq:Requation},\ref{eq:Sequation}) in its standard form, in which
$R_{,uv}$ and $S_{,uv}$ are explicitly given in terms of lower-order
derivatives: \begin{equation}
R_{,uv}=-e^{S}\frac{2R-K}{2\left(R-K\right)}-R_{,u}R_{,v}\frac{K}{2R\left(R-K\right)},\label{eq:RuvSTANDARD}\end{equation}
 \begin{equation}
S_{,uv}=e^{S}\frac{K}{2R\left(R-K\right)}+R_{,u}R_{,v}\frac{K}{2R^{2}\left(R-K\right)}.\label{eq:SuvSTANDARD}\end{equation}
 This form makes it obvious that the evolution equations become singular
when $R=K$, and also at $R=0$. This singularity was studied in some
detail in Ref. \cite{DO}.

\subsubsection*{Gauge freedom}

In a coordinate transformation $u \to  u'(u),v \to  v'(v)$
the dilaton scalar field $\phi$ is unchanged, but $\rho$ changes
as \begin{equation}
\rho'=\rho-\frac{1}{2}\left(\ln\frac{du'}{du}+\ln\frac{dv'}{dv}\right)\label{eq:RHOgauge}\end{equation}
(as may be deduces from the coordinate transformation of the metric
component $g_{uv}=-(1/2)e^{2\rho}$). From the definition (\ref{eq:RSdefinition})
of $R$ and $S$ it is obvious that $R$ is a scalar, and $S$ changes
in a coordinate transformation like $2\rho$:
\begin{equation}
R'=R\;,\;\;\; S'=S-\ln\frac{du'}{du}-\ln\frac{dv'}{dv}\:.\label{eq:Sgauge}\end{equation}

Below we shall often provide expressions for $S$ in certain specific
gauges. In such cases we shall use the notation $S_{[...]}$, with
the specific $u,v$ coordinates specified in the squared brackets.
The same notation will apply to $\rho$ (and also to the gauge-dependent
quantity $Z$ introduced in the next section).

\subsection{Classical solutions}

The classical solutions are obtained by setting $K=0$, leading to
$\hat{T}_{uu}=\hat{T}_{vv}=0$. The vacuum field equations then reduce
to the evolution equations \begin{equation}
R_{,uv}=-e^{S}\;,\;\;\; S_{,uv}=0\label{eq:EVOLUTIONclassic}\end{equation}
and the constraint equations\begin{equation}
R_{,uu}-R_{,u}S_{,u}=R_{,vv}-R_{,v}S_{,v}=0.\label{eq:CONSTRAINTclassic}\end{equation}
The general solution of these equations may be easily constructed.
It takes the form 
\begin{equation}
R(u,v)=M-R_{u}(u)R_{v}(v)\;,\;\;\;\; S(u,v)=\ln(R_{u,u}R_{v,v}),\label{eq:Rgencl}\end{equation}
where $M$ is an arbitrary constant, and $R_{u}(u)$ and $R_{v}(v)$
are any monotonically-increasing functions of their arguments (with
non-vanishing derivatives). The classical solution may thus look at
first glance as a rich class depending on two arbitrary functions.
However, these two functions merely reflect the gauge freedom. To
fix this freedom we may use the \emph{Kruskal-like} coordinates $U\equiv R_{u}(u),V\equiv R_{v}(v)$,
after which the general solution takes the simple explicit form \begin{equation}
R=M-UV\;,\;\;\; S_{[U,V]}=0.\label{eq:KRUSKALclassical}\end{equation}
The sub-index "$[U,V]$", recall, indicates that this expression
for $S$ only applies in a specific gauge, the one associated with
the Kruskal $U,V$ coordinates.

The representation (\ref{eq:KRUSKALclassical}) makes it obvious that
the classical vacuum solution is a one-parameter family, parametrized
by the mass $M$. We shall refer to it as the \emph{Schwarzschild-like}
solution. 

For $M>0$ the spacetime contains a BH, whose causal structure resembles
that of the four-dimensional Schwarzschild spacetime (see \cite{CGHS}).
The event horizon is located at $U=0$, and the past (or "white-hole")
horizon at $V=0$. Inside the BH ($U,V>0$) there is a spacelike $R=0$
singularity ($\phi,\rho$ diverge) 
at $UV=M$. For negative $M$ there is a naked, timelike,
$R=0$ singularity instead of a BH. 
(The $M=0$ case is considered below.) 

Another useful gauge is the \emph{Eddington-like} gauge, obtained
by the transformation 
\footnote{The coordinate $u_{e}$ introduced here is the \emph{classical} outgoing
Eddington coordinate. It should not be confused with the semiclassical
outgoing Eddington coordinate $\tilde{u}$ introduced in Sec. \ref{sec:Gauges}.
On the other hand the ingoing Eddington coordinate $v_{e}$ (denoted
later by $v$) is common to the classical and semiclassical solutions.} 
\begin{equation}
u_{e}\equiv-\ln(-U)\;,\;\;\; v_{e}\equiv\ln V\quad\quad(U<0,V>0).\label{eq:EDdef}\end{equation}
The solution then takes the form \begin{equation}
R=M+e^{v_{e}-u_{e}}\;,\;\;\; S_{[u_{e},v_{e}]}=v_{e}-u_{e}.\label{eq:EDclassic}\end{equation}
Note that these coordinates only cover the BH exterior. Analogous
Eddington-like coordinates (with appropriate sign changes) may also
be constructed for the BH interior, but we shall not need these internal
coordinates here. 

The special case $M=0$ yields the \emph{Minkowski-like} solution.
In Kruskal coordinates it takes the form \begin{equation}
R=-UV\;,\;\;\; S_{[U,V]}=0.\label{eq:KRUSKALM0}\end{equation}
In Eddington coordinates one obtains $R=e^{v_{e}-u_{e}},S_{[u_{e},v_{e}]}=v_{e}-u_{e}$
and hence $\rho_{[u_{e},v_{e}]}=0$. This demonstrates that the $M=0$
solution is \emph{flat}, and also indicates that the Eddington-like
coordinates $(u_{e},v_{e})$ correspond in this case to the standard,
flat, null coordinates of 2D Minkowski spacetime. 

In the general case $M\neq0$ one finds from Eqs. (\ref{eq:EDclassic}) and (\ref{eq:rho})
\begin{equation}
\rho_{[u_{e},v_{e}]}=-\frac{1}{2}\ln\left(1+\frac{M}{e^{v_{e}-u_{e}}}\right),\label{eq:RHOed}\end{equation}
and the associated metric $ds^{2}=-e^{2\rho}dudv$ yields non-vanishing
curvature. However, at large $v_{e}-u_{e}$ this expression reduces
to \begin{equation}
\rho_{[u_{e},v_{e}]}\cong-\frac{M}{2e^{v_{e}-u_{e}}}\cong-\frac{M}{2R},\label{eq:RHOedFAR}\end{equation}
which vanishes as $v_{e}-u_{e} \to \infty$. That is, in Eddington
coordinates $\rho$ vanishes at (right) PNI ($u_{e} \to -\infty$),
FNI ($v_{e} \to \infty$), and spacelike infinity ($u_{e} \to -\infty,v_{e} \to \infty$).
The Schwarzschild-like solution is thus \emph{asymptotically-flat},
and $(u_{e},v_{e})$ serve as asymptotically-flat coordinates in this
spacetime. In particular, $v_{e}$ and $u_{e}$ respectively coincide
with the affine parameter along PNI and FNI.

\subsubsection*{Classical collapsing shell}

The physical scenario which concerns us here is the formation of a
BH by the collapse of a thin shell, carrying a mass $M_{0}>0$, which
propagates along an ingoing null line $v=v_{0}$. The classical solution
describing this scenario is obtained by continuously matching a Minkowski-like
solution at $v\leq v_{0}$ with a Schwarzschild-like solution at $v\geq v_{0}$.
To describe the matching we start with the Kruskal-gauge solution
(\ref{eq:KRUSKALclassical}), along with its $M=0$ analog (\ref{eq:KRUSKALM0}).
Obviously the $R$-functions in these two expressions do not properly
match (at any $V$). To allow for continuous matching, we shift the
$U$ coordinate in the Minkowski-like domain $v<v_{0}$ , defining
$\hat{U}\equiv U+\Delta U$, where $\Delta U$ is a constant. Note that $S$ is unchanged in this
coordinate transformation. We obtain the Minkowski-like solution in
its new form:\begin{equation}
R=(\Delta U-\hat{U})V\;,\;\;\; S_{[\hat{U},V]}=0\quad\quad\quad\quad(v<v_{0}).\label{eq:KRUSKALflat}\end{equation}
This solution can now be matched to Eq. (\ref{eq:KRUSKALclassical}),
by equating $V\Delta U$ to $M_{0}$ at the null shell. Omitting
the hat from $\hat{U}$ and re-naming this new Minkowski-like Kruskal
coordinate by $U$, we obtain the matched solution
\begin{equation}
S_{[U,V]}=0\;,\;\;\; R(U,V)=\begin{cases}
(\Delta U-U)V\;\; & V<V_{0},\\
M_{0}-UV & V>V_{0},\end{cases}\label{eq:KRUSKALmatch}\end{equation}
where $V_{0}$ is the shell's $V$-value (we assume $V_{0}>0$), and continuity implies $\Delta U=M_{0}/V_{0}$.

To formulate the semiclassical initial data (next subsection) we find
it most convenient to work with Eddington $v$ and Kruskal $U$. The
collapsing-shell solution then takes the form 
\begin{equation}
S_{[U,v_{e}]}=v_{e}\;,\;\;\; R(U,v_{e})=\begin{cases}
(\Delta U-U)e^{v_{e}}\;\; & v_{e}<v_{e}^{0},\\
M_{0}-Ue^{v_{e}} & v_{e}>v_{e}^{0},\end{cases}\label{eq:EDDINGTONmatch}\end{equation}
where $v_{e}^{0}\equiv\ln V_{0}$.

\subsubsection*{Fixing the $v$-gauge}

Throughout the semiclassical analysis below, we find the Eddington
$v_{e}$ to be the most convenient $v$-coordinate, and no other $v$-coordinate
will be used. To simplify the notation, in the rest of this paper
we shall simply denote $v_{e}$ by $v$. 

For the $u$-gauge we shall occasionally use several $u$-coordinates
at various stages of the construction. The value of $S$ (like $\rho$
and $Z$) in specific $u$-gauges will be denoted by $S_{[...]}$---with
the specific $u$-coordinate only specified in the squared brackets---and
it will always refer to the Eddington $v$ coordinate. No confusion
should arise, because no $v$-coordinate other than Eddington will
be used throughout the rest of the paper. 

To practice the new notational rules we re-write the above classical
collapsing-shell solution (\ref{eq:EDDINGTONmatch}) in the new notation:
\begin{equation}
S_{[U]}=v\;,\;\;\; R(U,v)=\begin{cases}
(\Delta U-U)e^{v}\;\; & v<v_{0},\\
M_{0}-Ue^{v} & v>v_{0},\end{cases}\label{eq:EDDINGTONmatchNEW}\end{equation}
where $v_{0}=\ln V_{0}$ denotes the shell's location in Eddington-$v$,
and \[
\Delta U=M_{0}\, e^{-v_{0}}.\]

The collapsing-shell spacetime (\ref{eq:EDDINGTONmatchNEW}) is asymptotically-flat,
just like the pure Schwarzschild-like and Minkowski-like solutions.
Also, the Eddington coordinate $v$ serves as an affine parameter
at PNI. [This may be verified by switching to the Eddington coordinate
$u_{e}\equiv-\ln(-U)$ and noting that $\rho_{[u_{e}]}$ vanishes
at the PNI limit $u_{e} \to -\infty$.]

\subsection{Initial data for the semiclassical solution}

\subsubsection*{Initial conditions at the Shell}

We return now to the semiclassical problem of collapsing shell. The
semiclassical effects vanish at the portion $v<v_{0}$ of the collapsing-shell
spacetime, owing to its flatness.  
\footnote{Flatness means that the term $\rho_{,uv}$ in the semiclassical evolution equations (\ref{eq:Requation},\ref{eq:Sequation}) vanish, so these equations reduce to the classical ones.} 
Therefore, the portion $v<v_{0}$
is correctly described by Eq. (\ref{eq:EDDINGTONmatchNEW}) even in
the semiclassical problem. 

The portion $v>v_{0}$ of the collapsing-shell spacetime will be profoundly
modified by semiclassical effect (as expressed for example by the
BH evaporation). However, by continuity, the initial conditions for
$R$ and $S$ at $v=v_{0}$ will still be determined by matching to
the same, unmodified, Minkowski-like solution at $v<v_{0}$, and will
therefore be exactly the same as in the classical solution (\ref{eq:EDDINGTONmatchNEW}).
Using again the Kruskal-$U$ coordinate (and, recall, the Eddington-$v$
coordinate as usual), these shell initial conditions take the form
\begin{equation}
R_{0}(U)\equiv R(U,v_{0})=M_{0}-Ue^{v_{0}}\label{eq:R0U}\end{equation}
and\begin{equation}
S_{0}^{[U]}(U)\equiv S_{[U]}(U,v_{0})=v_{0}.\label{eq:S0U}\end{equation}

\subsubsection*{Initial conditions at Past null infinity}

The semiclassical evolution equations (\ref{eq:Requation},\ref{eq:Sequation})
also require initial conditions at an outgoing null ray, which
is most conveniently taken to be the PNI boundary. The situation here
is somewhat similar to that of the initial data at the shell: Owing
to asymptotic flatness of the collapsing-shell spacetime [and to
the lack of any strong-field region at the causal past of PNI (unlike
the situation at FNI)], the asymptotic behavior at PNI is just the
classical one. Specifically one finds the initial conditions\begin{equation}
R\cong M_{0}-Ue^{v}\;,\;\;\; S_{[U]}\cong v\quad\quad\quad\quad(PNI).\label{eq:PNIdata}\end{equation}
in the domain $v>v_{0}$. In particular, the influx $\hat{T}_{vv}$
should vanish at PNI (corresponding to an incoming vacuum state).

\subsection{The role of $K$ in semiclassical dynamics}

The semiclassical effects originate from the last term $K\rho_{,u}\rho_{,v}$
in the CGHS action (\ref{eq:Action}). The parameter $K=N/12$ thus
determines the overall magnitude of semiclassical effects. It is important
to note, however, that $K$ may be factored out from CGHS dynamics
by a simple shift/rescaling of variables. 
In the transformation 
\begin{equation}
N \to  cN\;,\;\;\; K \to  cK \;,\;\;\; \phi \to \phi-\frac{1}{2}\ln c
\label{eq:Kscale}
\end{equation}
[with $\lambda$, $\rho$ and $f_i$ unchanged (and with all scalar fields $f_i$ identical)], 
the action is merely multiplied by $c$, hence the field equations
are unaffected. 
One can therefore factor out $K$ this way by choosing $c=1/K$. 

In the relevant domain $v>v_{0}$ the scalar fields $f_{i}$ vanish anyway, 
so the term $-\frac{1}{2}\sum\limits _{i=1}^{N}f_{i,u}f_{i,v}$ is absent  
from the action (\ref{eq:Action}). 
In terms of the new variables $R,S$, the rescaling law takes the
form
\begin{equation}
K \to cK \;,\;\;\; R \to cR\;,\;\;\;\; S \to S+\ln c\;.\label{eq:KRscale}\end{equation}

This scaling makes it obvious that the relative magnitude of semiclassical
effects in various regions of spacetime will not be determined by
$K$ or $R$ separately, but only through (dimensionless 
\footnote{Note that after $\lambda$ is set to $1$ [see discussion following
Eq. (\ref{eq:Action})] all model's quantities become dimensionless.
This includes $R$, and also the BH mass $M_{0}$. (The
latter is naturally linked to $R$---for example, through the horizon's
$R$ value.)
}) combinations like $K/R$
or $K/e^S$, which are invariant to the rescaling. 
\footnote{It is important not to confuse here between two different issues related
to the magnitude of $K$: First, whether the semiclassical treatment
is applicable or not. Second, \emph{within the semiclassical formulation},
$K$ determines the magnitude of semiclassical effects. CGHS pointed
out \cite{CGHS} that the semiclassical treatment is only valid if
$K\gg1$. Here we address the second issue. Namely, we assume that
the condition $K\gg1$ is satisfied, and explore the scaling law which
characterizes the magnitude of semiclassical effects (and its relation
to the scale of other variables like $R$).}

One can easily verify that in the above scaling transformation the
BH original mass $M_{0}$ is multiplied by $c$. It is a common wisdom
that when a BH is "macroscopic", semiclassical effects will
be locally weak 
\footnote{Here, again, it is important to distinguish between two different
aspects of "macroscopicality": (i) Whether the semiclassical
theory is applicable to the BH evaporation; and (ii) whether, within
the semiclassical formulation, the semiclassical effects are locally
weak. The discussion here pertains to the second aspect. To satisfactorily
deal with (i) one has to further assume $K\gg1$ (see also previous
footnote). } 
(except in the neighborhood of the singularity). The above scaling
law makes it obvious that whether a BH may be regarded as "macroscopic"
or not, would only depend on the ratio of $M_{0}$ and $K$: A CGHS
BH should be regarded as macroscopic if $M_{0}\gg K$---which we indeed
assume throughout this paper.

In principle, the above scaling allows us to set $K=1$ in the field
equations. We find it more convenient, however, to leave $K$ in the
equations untouched. The parameter $K$ serves as a "flag" marking
the terms of semiclassical origin in the various equations. Also,
in the equations below $K$ always appears through combinations like
$K/R$, $K/R_{0}$, $K/M_{0}$ (or sometimes with $R$ replaced by
the variable $W\sim R$ introduced below). We shall assume throughout
this paper that the BH is macroscopic ($M_{0}\gg K$), and deal with
spacetime regions satisfying $R\gg K$. This will allow us to expand
various expressions to first order in small quantities such as $K/R$
or alike. All these expansions may conveniently be handled formally
as expansions in $K$ (or in the parameter $q=K/4$ introduced below)---though
one should bear in mind that the small parameter in the expansion is not $K$ itself, but the combinations $K/R$, $K/M_{0}$, etc.

\section{Field redefinition and flux-conserving formulation \label{sec:Redefinition}}

\subsection{Field redefinition}

The evolution equations (\ref{eq:Requation},\ref{eq:Sequation})
may look quite simple at first glance. However, to close the system
one must substitute Eq. (\ref{eq:rho}) for $\rho$, which makes the equations rather messy. Bringing these equations to their
standard form, one ends up with Eqs. (\ref{eq:RuvSTANDARD},\ref{eq:SuvSTANDARD}).
In addition to their rather complicated form, these second-order equations
also have the inconvenient property of being explicitly dependent on
first-order derivatives $R_{,u},R_{,v}$. 

To get rid of this undesired dependence upon $R_{,u}$ and $R_{,v}$, we
transform from $R$ and $S$ to new variables $W,Z$ defined by \cite{bilal}
\begin{equation}
W(R)\equiv\sqrt{R(R-K)}-K\ln(\sqrt{R}+\sqrt{R-K})+K\left(\frac{1}{2}+\ln2\right)\label{eq:Wdef}\end{equation}
and \begin{equation}
Z\equiv S+\Delta Z(R),\label{eq:Zdef}\end{equation}
where \begin{equation}
\Delta Z(R)\equiv\frac{2}{K}(R-W)-\ln R.\label{eq:dZdef}\end{equation}
With these new variables the evolution equations take the schematically-simpler
form: \begin{equation}
W_{,uv}=e^{Z}\, V_{W}(W)\;,\quad Z_{,uv}=e^{Z}\, V_{Z}(W),\label{eq:potentialForm}\end{equation}
with certain "potentials" $V_{W}(W)$ and $V_{Z}(W)$. But the
simplification does not come without a cost: These potentials are
explicitly obtained as functions of $R$ rather than $W$. One finds
\begin{equation}
V_{W}=-\frac{R-K/2}{\sqrt{R(R-K)}}\, e^{-\Delta Z(R)}\label{eq:VWdef}\end{equation}
 and \begin{equation}
V_{Z}=\frac{2}{K}\left[\frac{R-K/2}{\sqrt{R(R-K)}}-1\right]\, e^{-\Delta Z(R)}.\label{eq:VZdef}\end{equation}
Despite this disadvantage, the form (\ref{eq:potentialForm}) allows
an effective treatment of the semiclassical dynamics, as will be demonstrated
below.

Notice that the gauge transformation (\ref{eq:Sgauge}) of $R$ and
$S$ carries over to $W$ and $Z$ respectively: \begin{equation}
W'=W\:,\:\:\: Z'=Z-\ln\frac{du'}{du}-\ln\frac{dv'}{dv}\:.\label{eq:Zgauge}\end{equation}
The value of $Z$ in a specific $u$-gauge will be denoted by $Z_{[...]}$, in full analogy with our notation for $S$. 

\subsection{Large-$R$ asymptotic behavior}

From now on we shall consider a macroscopic BH, $M_{0}\gg K$, and
restrict the analysis to spacetime regions where $R\gg K$. We expand
Eqs. (\ref{eq:Wdef},\ref{eq:dZdef}) to first order in the small
quantity $K/R$:
\begin{equation}
W=R\left[1-\frac{K}{2R}\ln R+O\left(\frac{K}{R}\right)^{2}\right],
\end{equation}
 \begin{equation}
\Delta Z=-\frac{K}{4R}+O\left(\frac{K}{R}\right)^{2}.\end{equation}
 The inverse function $R(W)$ is given at this order by \begin{equation}
R=W\left[1+\frac{K}{2W}\ln W+O\left(\frac{K}{W}\right)^{2}\right].
\end{equation}
 The potentials $V_{W}(W),V_{Z}(W)$ can now easily be expanded to
first order in $K/W$, \begin{equation}
V_{W}=-1-\frac{K}{4W}+O\left(\frac{K}{W}\right)^{2}\,\,,\,\,\,\,\, V_{Z}=\frac{1}{W}\left[\frac{K}{4W}+O\left(\frac{K}{W}\right)^{2}\right].\label{eq:VZaprx}\end{equation}

\subsection{The flux-conserving system}

We now proceed with the above large-$R$ approximation, keeping only
terms up to first order in $K/R$ (or $K/W$); Thus we analyze
the hyperbolic system \begin{equation}
W_{,uv}=e^{Z}V_{W}(W)\;,\quad Z_{,uv}=e^{Z}V_{Z}(W)\label{eq:system}\end{equation}
 with \begin{equation}
V_{W}=-1-\frac{q}{W}\;,\quad V_{Z}=\frac{q}{W^{2}}\;,\label{eq:VWZ}\end{equation}
 where \[
q\equiv\frac{K}{4}.\]

Since $V_{Z}=dV_{W}/dW$, Eqs. (\ref{eq:system},\ref{eq:VWZ}) constitute
a \emph{flux-conserving system}. This concept was first introduced in Ref.
\cite{Ori-charged} and was later described in more detail in Refs. \cite{gorbonos}
and \cite{Ori-flux}. Our system corresponds to $F=V_{W}=-1-q/4$,
hence the generating function \cite{notation} is \begin{equation}
h_{0}(W)=W+q\ln W.\label{eq:h(W)}\end{equation}

In conjunction with the system (\ref{eq:system},\ref{eq:VWZ}) we shall also use the transformation between ($R,S$) and ($W,Z$) chopped at first order in $q$, namely 
\begin{equation}
W=R-2q\ln R,\label{eq:RtoW}\end{equation}
\begin{equation}
R=W+2q\ln W,\label{eq:WtoR}\end{equation}
\begin{equation}
Z=S-\frac{q}{R}=S-\frac{q}{W},\label{eq:StoZ}\end{equation}
\begin{equation}
S=Z+\frac{q}{R}=Z+\frac{q}{W}.\label{eq:ZtoS}\end{equation}
We also rewrite here the shell initial conditions (\ref{eq:R0U},\ref{eq:S0U}) in the new variables $W$ and $Z$, to first order in $q$:
\begin{equation}
W_{0}(U)\equiv W(U,v_{0})=R_{0}-2q\ln R_{0}\label{eq:W0U}\end{equation}
and\begin{equation}
Z_{0}^{[U]}(U)\equiv Z_{[U]}(U,v_{0})=v_{0}-\frac{q}{R_{0}}.\label{eq:Z0U}\end{equation}

\subsubsection*{Ingoing Vaidya-like solutions}

Like any flux-conserving system, the system (\ref{eq:system},\ref{eq:VWZ}) admits \emph{Vaidya-like solutions}---namely, solutions with a single flux
\cite{notation}. Each Vaidya-like solution is endowed with a \emph{mass-function}---a
function of one of the null coordinates which encodes the information
about the flux. As it turns out, at the leading order an evaporating
BH may be approximated by an \emph{ingoing} Vaidya-like solution with a linearly-decreasing
mass function. [However, a first-order outgoing component has to
be superposed on it in order to precisely match the initial conditions,
as we discuss below.] Assuming that the BH was created by the collapse
of a shell of mass $M_{0}$, which propagated along the null orbit
$v=v_{0}$, the appropriate mass function takes the form $\bar{M}(v)=M_{0}-q(v-v_{0})$.
This choice is motivated by the well-known fact \cite{CGHS} that
a 2D macroscopic BH evaporates at a constant rate $q=N/48$; and the
suitability of (the approximate solution derived from) this mass
function is verified in Sec. \ref{sec:Verification} below. By shifting
the origin of $v$ such that $v_{0}=-M_{0}/q$, we obtain the mass
function in a more compact form: \begin{equation}
\bar{M}(v)=-qv\equiv m_{v}(v).\label{eq:mv}\end{equation}

The restriction to a Vaidya-like solution leads to a great simplification,
because the problem now reduces to that of solving an ODE (rather
than a system of PDEs). Thus, as described in Ref. \cite{Ori-flux},
$W(u,v)$ is now determined from the ODE $W_{,v}=h_{0}(W)-\bar{M}(v)$
(applied along each line $u=const$), or, more explicitly,
\begin{equation}
W_{,v}=(W+q\ln W)+qv.\label{eq:Wv}\end{equation}
 The other unknown $Z(u,v)$ is then given by \begin{equation}
Z=\ln(-W_{,u}).\label{eq:Zv1}\end{equation}
Alternatively \cite{Ori-flux} $Z$ may be obtained from the ODE
 \begin{equation}
Z_{,v}=1+\frac{q}{W}.\label{eq:Zv}\end{equation}

As was mentioned above, we set the origin of $v$ such that the collapsing
shell is placed at $v_{0}=-M_{0}/q$. Therefore the parameter $v_{0}$
is negative. Also, throughout this paper we assume that the BH is
macroscopic, namely $M_{0}\gg q$, and this implies $v_{0}\ll-1$.

\subsubsection*{Weakly-perturbed Vaidya-like solution}

The above mentioned ingoing Vaidya-like solution well approximates
many aspects of the CGHS spacetime. However, it fails to precisely
match the initial conditions at the shell. This mismatch is small,
$\propto(q/W)$, yet it must be fixed in order to satisfactorily handle
some of the more subtle aspects of the solution (most importantly, the
Hawking outflux at FNI). Thus, we must fix the ingoing Vaidya-like
solution by adding to it a small outgoing component, seeded by the
mismatch at the shell. Nevertheless, owing to the small $\propto(q/W)$
magnitude of the mismatch, it will be possible to treat this outgoing
component as a small (linear) perturbation on top of the ingoing Vaidya-like
solution discussed above (to which we shall refer as the "core solution"). 

In the next section we shall proceed with analyzing the ingoing Vaidya-like
core solution. Then in Sec. \ref{sec:perturb} we shall construct
the perturbing outgoing component and thereby complete the construction
of the approximate solution.

\section{Constructing the ingoing Vaidya-like core solution \label{sec:Construction}}

\subsection{Processing the ODE for $W$ and introducing $H$ \label{sec:ODEforH}}

The term $q\ln W$ in the right-hand side of Eq. (\ref{eq:Wv}) makes
this equation hard to analyze. In order to ease the analysis we define
the auxiliary variable $H\equiv W_{,v}+q$, namely \begin{equation}
H=W+q\ln W-m_{v}+q.\label{eq:Hdef}\end{equation}
The inverse function $W(H)$ cannot be expressed in a closed exact
form; However, restricting the analysis to first order in $q$ we
may use the relation\begin{equation}
W=H+m_{v}-q[\ln(H+m_{v})+1].\label{eq:W1}\end{equation}

Differentiating now $H$ using Eq. (\ref{eq:Hdef}) and $W_{,v}=H-q$
we find\[
H_{,v}=q+\frac{\partial H}{\partial W}W_{,v}=q+\left(1+\frac{q}{W}\right)(H-q).\]
Further substituting Eq. (\ref{eq:W1}) in the right-hand side and
omitting all $\propto q^{2}$ terms, we obtain the ODE for $H$ in
its more compact form: \begin{equation}
H_{,v}=H\left(1+\frac{q}{H-qv}\right).\label{eq:Hdv}\end{equation}

The initial conditions for $H$ is to be specified at the shell's orbit,
the line $v=v_{0}$ (this line serves as a characteristic initial
surface for the non-trivial piece $v>v_{0}$ of the CGHS spacetime).
We denote it $H_{0}(u)\equiv H(u,v_{0})$. In Sec. \ref{sub:Calculating-H0}
we calculate $H_{0}$ and show that it decreases linearly with the
Kruskal coordinate $U$. Then in Sec. \ref{sec:Gauges} we express
$H_{0}(u)$ in a few other useful gauges. Note that the dependence
of $H$ on $u$ only emerges through the initial condition $H_{0}(u)$.
Correspondingly we shall often express this parametric dependence
in the form $H(v;u)$, and sometimes ignore it altogether and use the abbreviated notation $H(v)$, for convenience. 

We are unable to solve the ODE (\ref{eq:Hdv}) analytically. However,
an approximate solution is given in Appendix \ref{sec:Appendix-H}.
Furthermore, several exact key properties of the solutions of Eq.
(\ref{eq:Hdv}) are described below.

Note the advantage of the ODE of $H$ over Eq. (\ref{eq:Wv}) for
$W_{,v}$, particularly at large $H$. The latter
corresponds to future or past null infinity, where, as it turns out,
$H$ (like $W$) grows exponentially in $v$. The ODE then reduces
to the trivial one $H_{,v}\cong H+q$, which is easily solved (leading
to the above-mentioned exponential growth). In the ODE for $W$, on
the contrary, one faces the $\ln W$ term, which complicates the asymptotic
behavior at future/past null infinity.

\subsection{Some exact properties of  $H(v;u)$ \label{sub:Properties}}

Given the ODE (\ref{eq:Hdv}), the function $H(v;u)$ is uniquely
determined by the initial-value function $H_{0}(u)\equiv H(u,v_{0})$
[this function is explicitly constructed below; cf. Eqs. (\ref{eq:H0U},\ref{eq:H0})].
We therefore start our discussion here by mentioning two key properties
of $H_{0}(u)$ which are important for the present analysis: First,
$H_{0}(u)$ is monotonically decreasing. It is positive at early $u$
but becomes negative afterwards. It vanishes at a certain $u$ value
which we denote $u^{hor}$ (in Kruskal gauge it corresponds to $U=-qe^{-v_{0}}$).
Second, $H_{0}(u)+M_{0}$ is positive at any $u$. 

We first note that the ODE (\ref{eq:Hdv}) admits a trivial solution
$H(v)=0$. It immediately follows that $H$ vanishes along the line
$u=u^{hor}$ (but nowhere else). 

Next, we define (off the line $u=u^{hor}$ where $H$ vanishes)\begin{equation}
l(v;u)\equiv\ln|H(v;u)|.\label{eq:ldef}\end{equation}
 It satisfies the ODE
\begin{equation}
l_{,v}=1+\frac{q}{H+m_{v}}=1+\frac{q}{\pm e^{l}-qv},\label{eq:eql}\end{equation}
where the "$\pm$" sign reflects the sign of $H$.

Both equations (\ref{eq:eql}) and (\ref{eq:Hdv}) develop a singularity
whenever $H+m_{v}$ vanishes, but are regular otherwise. Since $H_{0}+M_{0}>0$,
the quantity $H+m_{v}$ is always positive at $v=v_{0}$ and its neighborhood.
In principle there could be two possibilities, which may depend on
$u$ [through the initial value $H_{0}(u)$]: (i) $H(v;u)$ is
regular throughout $v>v_{0}$, or (ii) $H(v;u)$ becomes singular
($H+m_{v}$ vanishes) at a certain finite $v=v_{sing}(u)>v_{0}$.
\footnote{As will become obvious later, option (i) occurs at $u<u^{hor}$ and
option (ii) at $u>u^{hor}$.} 
To treat both cases in a unified manner we shall say that the solution
$H(v;u)$ is regular and well-defined throughout the domain $v_{0}\leq v<v_{f}(u)$, where $v_{f}(u)=\infty$ in case (i) and $v_{f}(u)=v_{sing}(u)$ in
case (ii). 

We can now deduce the following exact properties of $H(v;u)$, which
hold throughout the domain of regularity $v<v_{f}(u)$: 

(a) As was already mentioned above, $H$ vanishes along the line $u=u^{hor}$.
This holds throughout the range $v_{0}\leq v<0$. [At $(u=u^{hor},v \to 0)$
the ODE becomes singular because $H+m_{v}$ vanishes.] $H$ cannot
vanish at any other value of $u$, because otherwise $H_{0}(u)$ would
have to vanish too at that specific $u\neq u^{hor}$ (which is not
the case). 

(b) Consequently, $H(u,v)$ has the same sign as $H_{0}(u)$ at any
$u$. As was mentioned above, this sign is positive at $u<u^{hor}$
and negative at $u>u^{hor}$. These two domains are separated by the
line $u=u^{hor}$ on which $u$ vanishes. 

(c) Since $H+m_{v}$ is positive at $v=v_{0}$, it remains positive
throughout $v_{0}<v<v_{f}(u)$. 

(d) Correspondingly, the right-hand side in Eq. (\ref{eq:eql}) is
strictly positive, and in fact $>1$. Thus, $l(v)$ is monotonically
increasing, and the same for $|H(v)|$. 

(e) The quantity $H_{,u}$ satisfies the ODE 
\[
\frac{d}{dv}\ln\left|H_{,u}\right|=1-\frac{q^{2}v}{(H+m_{v})^{2}}\]
[cf. Eq. (\ref{eq:HduDIF}) below]. The right-hand side is regular
at any $v<v_{f}(u)$, and positive throughout $v\leq0$. Since $H_{0,u}$
is negative for all $u$, 
\footnote{$H_{0}$ decreases linearly with $U$; and we only consider here $u$-gauges satisfying $du/dU>0$.} 
it immediately follows from this ODE that $H_{,u}$ is everywhere
negative. Furthermore, at least at $v<0$, $\left|H_{,u}\right|$
is an increasing function of $v$; therefore, at fixed $u$ (and $v<0$)
$H_{,u}$ is bounded above by the parameter $H_{0,u}<0$. 

(f) The same obviously applies to $(H+m_{v})_{,u}$ (in particular,
this quantity is everywhere negative). It then follows that if a singularity
$H+m_{v}=0$ occurs at some $u=u_{1}$ (with finite $v_{sing}>v_{0}$),
then such a singularity must occur at any $u>u_{1}$, and $v_{sing}(u)$
must be a non-increasing function of $u$ (throughout $u>u_{1}$).
Furthermore, in the range where $v_{sing}(u)<0$ [see (g)], $v_{sing}(u)$
must be a \emph{strictly-decreasing} function of $u$.

(g) As was mentioned above, at $u>u^{hor}$ $H_{0}(u)$ is negative,
and the same for $H(v;u)$. [Furthermore, since $|H(v)|$ is monotonically
increasing, at each line $u=const$ in this domain $H$ is bounded
above by the parameter $H_{0}(u)<0$.] It then follows that $H+m_{v}$
(which is positive at $v=v_{0})$ must vanish before $m_{v}$ vanishes.
Thus, all lines $u>u^{hor}$ run into an $H+m_{v}=0$ singularity
at a certain $v=v_{sing}(u)<0$. Property (f) then implies that $v_{sing}$
is a strictly-decreasing function of $u$, indicating that this is
a \emph{spacelike} singularity.

(h) In the domain $u<u^{hor}$, $H$ is strictly positive, and no
singularity may form at $v\leq0$ (where $m_{v}$ is positive too).
However, at $v>0$ $m_{v}$ is negative, and one might be concerned
about the possibility of vanishing $H+m_{v}$, which would lead to
a singularity. A closer examination reveals that such a singularity
does not occur in this domain. This may be deduced from each of the
following arguments (though a complete mathematical proof is still
lacking): (i) It is easy to show that at least a locally-monotonic
singularity of this type (namely, a singularity of vanishing $H+m_{v}$
at some finite $v=v_{sing}$, such that $H+m_{v}$ is monotonic in
$v$ throughout some neighborhood of $v=v_{sing}$) is not possible
in the domain $u<u^{hor}$: $H+m_{v}$ starts positive at $v=v_{0}$,
and in order to vanish it must \emph{decrease} on approaching $v=v_{sing}$.
However, since $H>0$ (and increasing), when $H+m_{v}$ approaches
zero from above Eq. (\ref{eq:Hdv}) yields $H_{,v} \to +\infty$,
and therefore $H+m_{v}$ must \emph{increase}, so it cannot vanish.
It is harder to mathematically exclude the possibility of an oscillatory
approach to an $H+m_{v}=0$ singularity, but such an oscillatory behavior
seems very unlikely. (ii) Consider the limiting function $H(v;u_{-}^{hor})$,
defined to be the limit $u \to  u_{-}^{hor}$ of $H(v;u)$.
At $v<0$ this function vanishes, just like $H(v;u^{hor})$, by continuity.
At $v\geq0$ [where $H(v;u^{hor})$ is not defined; see (a)],
this function becomes a non-trivial solution of the ODE (\ref{eq:Hdv}).
Numerical examination shows that $H(v;u_{-}^{hor})$ continuously
increases from zero at $v=0$ to infinity at $v \to \infty$,
with $H+m_{v}>0$ at any $v>0$. From (f) it now follows that $H+m_{v}>0$
throughout $u<u^{hor}$. (iii) Direct numerical simulations of $H(v;u)$
at various $u<u^{hor}$ values further confirm this conclusion. 

Summarizing the above discussion on the exact properties of $H(u,v)$,
and briefly re-stating it in more physical/geometrical terms: All
lines $u<u^{hor}$ make it to FNI, whereas all lines
$u>u^{hor}$ run into a spacelike singularity 
\footnote{The \emph{exact} semiclassical CGHS spacetime admits a very similar global structure: A BH, with a spacelike singularity inside it. We
point out, however, that the \emph{local} properties of the spacelike
singularity in our approximate solution are different from those of
the precise CGHS spacetime. In particular, in our approximate solution
(when literally applied to the $H+m_{v}=0$ singularity) $R$ diverges
logarithmically to $-\infty$, whereas in the CGHS solution $R=K$
at the singularity \cite{DO}. This remarkable difference is not a
surprise, because the parameter $q/R$ is no longer small when we
get close to the singularity, hence our approximate solution becomes
invalid there. } 
at $v=v_{sing}(u)<0$. 

Based on these causal properties of $H(u,v)$ and its singularity,
we shall refer to the ranges $u<u^{hor}$ and $u>u^{hor}$ as the
BH exterior and interior, respectively. Note that the interior is
confined to the range $v_{0}\le v<v_{sing}(u)<0$ (and $u>u^{hor}$), whereas the exterior extends in the entire domain $v_{0}\le v<\infty$ (for $u<u^{hor}$).

\subsection{Expressing the ingoing solution in terms of $H$}

The function $H(u,v)$ may be obtained by solving the ODE (\ref{eq:Hdv})
numerically, or by analytic approximate solutions like the one given
in Appendix \ref{sec:Appendix-H}. Once $H$ is known, $W$ is given
by Eq. (\ref{eq:W1}), and $Z$ in turn by Eq. (\ref{eq:Zv1}). The
former equation yields \[
W_{,u}=(1-\frac{q}{H+m_{v}})H_{,u},\]
hence, up to first order in $q$, \begin{equation}
Z=\ln(-H_{,u})-\frac{q}{H+m_{v}}.\label{eq:ZinH}\end{equation}
One can easily verify the consistency of this expression with the
$u$-gauge transformation of $Z$, given in Eq. (\ref{eq:Zgauge})
(note that $H$ is unchanged in such a transformation).

\subsection{Calculating $H_{0}(u)$ \label{sub:Calculating-H0}}

To calculate $H_{0}(u)$ we evaluate Eq. (\ref{eq:ZinH}) at $v=v_{0}$
and equate it to the desired initial condition $Z_{0}(u)$. It is
convenient to carry out this calculation in the Kruskal gauge. Using
Eqs. (\ref{eq:Z0U},\ref{eq:R0U}) we obtain the following equation
for $H_{0}$: \begin{equation}
\ln(-H_{0,U})-\frac{q}{H_{0}+M_{0}}=v_{0}-\frac{q}{M_{0}-Ue^{v_{0}}}.\label{eq:H0eq}\end{equation}
A simple solution immediately suggests itself: $H_{0}=-Ue^{v_{0}}$;
but we need here the general solution for this ODE, which spans a
one-parameter family. For our purpose it will be sufficient, however,
to derive an \emph{approximate} general solution (up to order $q$),
which is an easy task. Such a one-parameter family of approximate
solutions is\begin{equation}
H_{0}(U)=-Ue^{v_{0}}+pq,\label{eq:H0gen}\end{equation}
where $p$ is a yet-arbitrary constant. This constant may be fixed
by solving the ODE (\ref{eq:Hdv}) for $H(v;U)$ in the PNI asymptotic
limit [with the above expression for $H_{0}(U)$ as initial condition],
constructing $W,Z$ from $H$ and then $R,S$, and comparing them
to the desired initial data at PNI. In Appendix \ref{sec:Appendix-Const}
we carry out this analysis and find that the appropriate value is
$p=-1$, namely\begin{equation}
H_{0}(U)=-Ue^{v_{0}}-q.\label{eq:H0U}\end{equation}
We may also use Eq. (\ref{eq:R0U}) to re-write $H_{0}$ in a form
which is explicitly gauge-invariant (for arbitrary $u$ coordinate):
\begin{equation}
H_{0}(u)=R_{0}(u)-M_{0}-q.\label{eq:H0}\end{equation}

\section{The weak perturbing outflux \label{sec:perturb}}

The above ingoing Vaidya-like solution was constructed (through an
appropriate choice of $H_{0}(u)$) such that it properly matches the
initial function $Z_{0}(u)$ at the shell. However, the other initial
function $W_{0}(u)$ does not exactly coincide with $W$ of the above
constructed ingoing solution. To fix this mismatch we shall add a
weak, outgoing component as a perturbation on top of the ingoing core
solution. 

To this end we write the overall $W,Z$ functions as \begin{equation}
W(u,v)=W^{in}(u,v)+\delta W(u,v),\label{eq:W2parts}\end{equation}
\begin{equation}
Z(u,v)=Z^{in}(u,v)+\delta Z(u,v),\label{eq:Z2parts}\end{equation}
where $W^{in},Z^{in}$ denote the Vaidya-like ingoing core solution 
constructed in the previous section 
[namely Eqs. (\ref{eq:W1}) and (\ref{eq:ZinH})],
and $\delta W,\delta Z$ denote the additional perturbing component. 
\footnote{Note that in a coordinate transformation $Z^{in}$ transforms like $Z$, hence $\delta Z$ is invariant.}

We first calculate the mismatch in the initial condition for $W$. Equations
(\ref{eq:W1}) and (\ref{eq:H0U}) yield for $W^{in}$ at the shell 
\begin{equation}
W_{0}^{in}(U)=M_{0}-Ue^{v_{0}}-q[\ln(M_{0}-Ue^{v_{0}})+2]\label{eq:Win0}\end{equation}
(we have omitted the $q$ in the log argument, being a higher-order
term; and we do this occasionally in the equations below). This is
to be compared to the \emph{actual} initial data for $W$, Eq. (\ref{eq:W0U}),
which, together with Eq. (\ref{eq:R0U}), reads \begin{equation}
W_{0}(U)=M_{0}-Ue^{v_{0}}-2q\ln(M_{0}-Ue^{v_{0}}).\label{eq:W0Uexp}\end{equation}
The difference is thus\begin{equation}
\delta W_{0}(U)=-q[\ln(M_{0}-Ue^{v_{0}})-2].\label{eq:dW0}\end{equation}

Note that no mismatch is present in the shell data for $Z$, because
we have chosen $H_{0}(U)$ in the first place so as to properly match
$Z_{0}(U)$; therefore, \begin{equation}
\delta Z_{0}(U)=0.\label{eq:dZ0}\end{equation}

Next we substitute Eqs. (\ref{eq:W2parts},\ref{eq:Z2parts}) in the
full flux-conserving system (\ref{eq:system}), to obtain field equations
for $\delta W,\delta Z$. Since we are only interested in the solution
up to first order in $q$, and the mismatch initial data (\ref{eq:dW0})
are already $O(q)$, we may treat $\delta W,\delta Z$ as small perturbations,
satisfying the linearized equations \begin{equation}
\delta W_{,uv}=e^{Z}\left[V_{W,W}\delta W+V_{W}\delta Z\right],\label{eq:deltaWduv}\end{equation}
\begin{equation}
\delta Z_{,uv}=e^{Z}\left[V_{Z,W}\delta W+V_{Z}\delta Z\right].\label{eq:deltaZduv}\end{equation}
Furthermore, we only need to consider here the coefficients ($V_{W},V_{W,W},V_{Z},V_{Z,W}$)
at zero order in $q$---namely, $V_{W}=-1$ and $V_{W,W}=V_{Z}=V_{Z,W}=0$,
yielding the trivial system \begin{equation}
\delta W_{,uv}=-e^{Z}\delta Z\;,\quad\delta Z_{,uv}=0.\label{eq:EQdeltaWZ}\end{equation}
The initial conditions are Eqs. (\ref{eq:dW0},\ref{eq:dZ0}) at the
shell, and no contribution from PNI. 
\footnote{Since the perturbation we add here is an outgoing component, it is
assumed to be seeded at the shell only (any ingoing component would
be absorbed in the ingoing core solution in the first place). } 
For $\delta Z$ we immediately obtain 
\begin{equation}
\delta Z(u,v)=0.
\end{equation}
 In turn $\delta W$ satisfies the trivial equation $\delta W_{,uv}=0$,
yielding \[
\delta W(U,v)=\delta W_{0}(U)=-q[\ln(M_{0}-Ue^{v_{0}})-2].\]
Thus, the overall solution is\begin{equation}
W(U,v)=W^{in}(U,v)+\delta W_{0}(U)=H+m_{v}-q[\ln(H+m_{v})+\ln(M_{0}-Ue^{v_{0}})-1]\label{eq:W2partsEXP}\end{equation}
and
\begin{equation}
Z(u,v)=Z^{in}(u,v)=\ln(-H_{,u})-\frac{q}{H+m_{v}}.\label{eq:Z2partsGEN}\end{equation}
Finally we re-write Eq. (\ref{eq:W2partsEXP}) in a form which
includes no specific reference to the Kruskal coordinate $U$, by
replacing $M_{0}-Ue^{v_{0}}$ (in the log argument) with $H_{0}+M_{0}$,
using Eq. (\ref{eq:H0U}):
\begin{equation}
W(u,v)=H+m_{v}-q[\ln(H+m_{v})+\ln(H_{0}+M_{0})-1].\label{eq:W2partsGEN}\end{equation}

\section{The final approximate solution \label{sec:final-solution}}

Transforming the above results (\ref{eq:W2partsGEN},\ref{eq:Z2partsGEN})
from $W,Z$ to the original variables $R,S$, using Eqs. (\ref{eq:WtoR},\ref{eq:ZtoS}),
we obtain our approximate solution in its final form:\begin{equation}
R(U,v)=H+m_{v}+q[\ln(H+m_{v})-\ln(H_{0}+M_{0})+1]\label{eq:Rfinal}\end{equation}
and\begin{equation}
S(u,v)=\ln(-H_{,u}),\label{eq:Sfinal}\end{equation}
where $q=K/4$. The function $H(u,v)$, recall, is determined by the
ODE 
\footnote{Note that the parameter $q$ may easily be factored out of this ODE:
Defining $\tilde{H}\equiv H/q$, we obtain the ODE in its universal form 
$\tilde{H},v=\tilde{H}[1+1/(\tilde{H}-v)]$. }
\begin{equation}
H_{,v}=H\left(1+\frac{q}{H-qv}\right),\label{eq:HdvSUM}\end{equation}
with initial conditions $H_{0}(u)\equiv H(u,v_{0})$ given by\begin{equation}
H_{0}(u)=R_{0}(u)-M_{0}-q,\label{eq:H0SUM}\end{equation}
or, more explicitly (in the Kruskal gauge)\begin{equation}
H_{0}(U)=-Ue^{v_{0}}-q.\label{eq:H0Usum}\end{equation}
Approximate analytic expressions for $H(u,v)$ are given in Appendix
\ref{sec:Appendix-H}, cf. Eqs. (\ref{eq:h1},\ref{eq:h1ed}).

Note that this approximate solution \emph{precisely} matches the required
initial conditions at the shell, namely (using Kruskal $U$-gauge)
$R=M_{0}-Ue^{v_{0}}$ and $S_{[U]}=\ln(-H_{,U})=v_{0}$.

The expression (\ref{eq:Sfinal}) for $S$ requires $H_{,u}$. The
approximate analytic expressions (\ref{eq:h1},\ref{eq:h1ed}) can
be directly differentiated to yield $H_{,u}$. However, when $H(v;u)$
is obtained by numerically solving the ODE (\ref{eq:HdvSUM}), a direct
numerical $u$-differentiation may be inconvenient. In this case it
is easier to obtain $H_{,u}$ by numerically integrating the ODE it
satisfies:\begin{equation}
\frac{d}{dv}(H_{,u})=\left[1-\frac{q^{2}v}{(H-qv)^{2}}\right]H_{,u}\label{eq:HduDIF}\end{equation}
(this can be done simultaneously with the numerical integration of
the ODE of $H$ itself). The initial condition at the shell is obviously 
\begin{equation}
H_{,u}(u,v_{0})=\frac{d}{du}H_{0}(u).\label{eq:HduINIT}\end{equation}

\subsection{Alternative approximate expression for $S$}

As was already mentioned in Sec. \ref{sec:Redefinition}, the $Z$
function in the ingoing Vaidya-like solution can be obtained by either
of the equations (\ref{eq:Zv1}) or (\ref{eq:Zv}). The above analysis was
based on the former equation, and it led to the expression (\ref{eq:Sfinal}).
If one uses Eq. (\ref{eq:Zv}) instead, one can derive an alternative
expression for $S$: \begin{equation}
S^{alt}(u,v)=v_{0}+\ln\frac{H}{H_{0}(u)}+q\left[\frac{1}{H+m_{v}}-\frac{1}{H_{0}(u)+M_{0}}\right]-\ln\frac{du}{dU}.\label{eq:Salt}\end{equation}
This expression for $S$ looks more complicated, but it has the advantage
that it does not require $H_{,u}$. 

It should be emphasized that the two expressions (\ref{eq:Sfinal},\ref{eq:Salt})
for $S$ are \emph{not} exactly identical. Yet the difference appears
to be compatible with the anticipated error characterizing the entire
approximation scheme used here. At the same time we also point out
that so far the error in the approximation (\ref{eq:Salt}) for $S$
has not been explored as thoroughly as that in the original approximation
(\ref{eq:Sfinal}) (cf. Sec. \ref{sec:Verification}).

\section{Some useful gauges \label{sec:Gauges}}

Our approximate solution was presented in Eqs. (\ref{eq:Rfinal}-\ref{eq:H0SUM})
in a fully ($u$-)gauge-covariant form. The only reference to a specific
gauge was made in Eq. (\ref{eq:H0Usum}), which explicitly gave $H_{0}$
in terms of the Kruskal $U$-coordinate. 

In this section we shall introduce a few additional useful $u$-gauges:
The shifted-Kruskal gauge, and the (external as well as internal)
semiclassical Eddington gauge. These new gauges slightly simplify
the functional form of $H_{0}(u)$. More importantly, they are better
adopted to the global structure of the evaporating-BH spacetime (e.g.
the location of the horizon). 

Note that in all these gauges, we use for the $v$-gauge the same
Eddington coordinate $v$ (originally denoted $v_{e}$), as we do
throughout this paper.

\subsection{Shifted-Kruskal gauge ($\tilde{U}$)}

We define the shifted Kruskal coordinate \begin{equation}
\tilde{U}\equiv U+qe^{-v_{0}}.\label{eq:KruskalShifted}\end{equation}
Since this is a constant shift, $S$ is unchanged: $S_{[\tilde{U}]}=S_{[U]}$. 

The expression for $H_{0}$ slightly simplifies in this gauge:\begin{equation}
H_{0}(\tilde{U})=-\tilde{U}e^{v_{0}}.\label{eq:H0(shiftedU)}\end{equation}
Note that $\tilde{U}^{hor}=0.$

We point out that the solution in $(\tilde{U},v)$ coordinates [just
like in $(U,v)$ coordinates] covers the entire BH spacetime. As
may be obvious from the discussion in subsection \ref{sub:Properties},
the BH exterior and interior correspond to $\tilde{U}<0$ and $\tilde{U}>0$,
respectively.

\subsection{Semiclassical Eddington gauge ($\tilde{u}$)}

In the range $\tilde{U}<0$ (the BH exterior) we define the Semiclassical
Eddington coordinate $\tilde{u}$ by\begin{equation}
\tilde{u}\equiv-\ln(-\tilde{U})\quad\quad\quad(\tilde{U}<0).\label{eq:utilde}\end{equation}
$S$ is modified by this transformation according to $S_{[\tilde{u}]}=S_{[U]}-\tilde{u}$.
The initial function for $H$ now reads\begin{equation}
H_{0}(\tilde{u})=e^{v_{0}-\tilde{u}}.\label{eq:H0edditngton}\end{equation}

In the Semiclassical Eddington gauge the alternative expression (\ref{eq:Salt})
for $S$ reduces to the simpler form
\begin{equation}
S_{[\tilde{u}]}^{alt}(\tilde{u},v)=\ln H+q\left[\frac{1}{H+m_{v}}-\frac{1}{H_{0}+M_{0}}\right].\label{eq:SaltEDext}\end{equation}

Note that the coordinate $\tilde{u}$ only covers the BH exterior.

\subsection{Internal semiclassical Eddington gauge ($\bar{u}$)}

In the range $\tilde{U}>0$ (the BH interior) we define the internal
semiclassical Eddington coordinate $\bar{u}$ by\begin{equation}
\bar{u}\equiv\ln(\tilde{U})\quad\quad\quad(\tilde{U}>0).\label{eq:utildeInterior}\end{equation}
Now $S$ is modified according to $S_{[\bar{u}]}=S_{[U]}+\bar{u}$.
The initial function for $H$ takes the form $H_{0}(\bar{u})=e^{v_{0}+\bar{u}}$.

\section{Verification of the approximate solution \label{sec:Verification}}

Our approximate solution (\ref{eq:Rfinal}-\ref{eq:H0SUM}) was constructed
here through a rather indirect process, which involved the
transformation to new field variables $W,Z$, the large-$R$ approximation,
the formalism of flux-conserving systems, and their Vaidya-like
solutions. It is therefore important to directly examine the validity
of the resultant expressions.

Naturally this examination involves two independent parts: (i) checking
compliance with the field equations; and (ii) checking compatibility
with initial conditions, both at the shell ($v=v_{0}$) and at PNI. These two parts will be carried out in the next two
subsections.

\subsection{Compliance with the field equations (error estimate)}

To define the local error in the evolution equations we substitute
the approximate expressions (\ref{eq:Rfinal},\ref{eq:Sfinal}) in
the field equations (\ref{eq:RuvSTANDARD},\ref{eq:SuvSTANDARD})
and evaluate the error---namely, the deviation of $R_{,uv}$ and $S_{,uv}$
from their respective values (specified at the right-hand side of
these two equations). The error defined in this way is obviously gauge-dependent,
and we find it convenient to employ the semiclassical Eddington coordinate
$\tilde{u}$ (along with Eddington $v$) for this task. Using the
MATHEMATICA software we find that the local errors in the two equations
indeed scale as $q^{2}$, as anticipated. More specifically, the errors
scale as $R(q/R)^{2}$ for $R_{,\tilde{u}v}$ and as $(q/R)^{2}$
for $S_{,\tilde{u}v}$---both multiplied by certain functions of $m_{v}/R$.
This local error estimate applies to typical off-horizon strong-field
regions, namely, regions for which $m_{v}/R$ is of order unity but
not too close to $1$. The error decays exponentially in $|v-\tilde{u}|$
both at weak-field regions ($m_{v}/R \to 0$, corresponding
to $v-\tilde{u} \to \infty$) and near-horizon regions ($m_{v}/R \to 1$,
corresponding to $v-\tilde{u} \to -\infty$). 

The global, accumulated, long-term error is harder to analyze.
It may be evaluated by comparing the approximate solution (\ref{eq:Rfinal},\ref{eq:Sfinal})
to numerical simulations, but this is beyond the scope of the present
paper. In the next section, however, we shall evaluate the accumulated
error in both $R$ and $S$ in the neighborhood of the horizon ($\tilde{u} \to \infty$).

The error in the constraint equations (\ref{eq:Ruu},\ref{eq:Rvv})
is also found to be proportional to $q^{2}$, as may be expected (based
on the mutual consistency of the evolution and constraint equations).
However, the functional dependence of the pre-factor on $m_{v}/R$
is more subtle and will not be addressed here.

\subsection{Compatibility with the initial conditions}

\subsubsection*{Initial data at the shell}

At $v=v_{0}$ the expressions (\ref{eq:Rfinal}) and (\ref{eq:Sfinal})
reduce to $R=H_{0}+M_{0}+q$ and $S=\ln(-H_{0,u})$ respectively.
By virtue of Eq. (\ref{eq:H0Usum}), one obtains (using the Kruskal $U$ coordinate)
$R(U,v_{0})=M_{0}-Ue^{v_{0}}$ and $S_{[U]}(U,v_{0})=v_{0}$, which
exactly match the desired initial data (\ref{eq:R0U},\ref{eq:S0U}).
\footnote{Obviously, this also implies exact matching of $R$ and $S$ to the Minkowski-like solution at $v \le v_{0}$ in any other gauge.}

\subsubsection*{Initial data at past null infinity}

In Sec. \ref{sec:PFNI} we analyze the asymptotic behavior of our
approximate $R$ and $S$ at PNI, and verify that they
satisfy the required asymptotic behavior (\ref{eq:PNIdata}).
We also show that the influx $T_{vv}=R_{,v}S_{,v}-R_{,vv}$ vanishes
at PNI, as it should.

\section{Horizon \label{sec:Horizon}}

In Sec. \ref{sec:Construction} we analyzed the behavior of $H(u,v)$
along $u=const$ lines, and found that spacetime is divided into two
domains by a certain outgoing null ray which (for a general $u$-gauge)
we denoted $u=u^{hor}$: All lines $u<u^{hor}$ run to FNI in a regular manner (with steadily growing $H$), whereas
all lines $u>u^{hor}$ crush into a spacelike singularity. The spacetime
thus contains a BH, and the domains $u<u^{hor}$ and $u>u^{hor}$
correspond to the BH exterior and interior, respectively. We shall
therefore regard the critical null ray $u=u^{hor}$ as the \emph{event
horizon} (or sometimes just \emph{horizon}) of the BH .
\footnote{This involves some abuse of standard terminology, because the line
$u=u^{hor}$ actually contains a naked singularity at $v=0$ 
(the point denoted "P" in Fig. 1), 
where $H+m_{v}$ vanishes. Note that it is only the section $v<0$ of this
line which separates the BH interior and exterior. The portion $v>0$
of $u=u^{hor}$ is in fact a \emph{Cauchy horizon} (see Fig. 1). }

The horizon is characterized by the vanishing of $H$, hence its location
$u=u^{hor}$ is determined by requiring $H_{0}(u)=0$. Referring to
some specific gauges, the horizon's location is $\tilde{U}=0$ in
the shifted Kruskal gauge, and $U=-qe^{-v_{0}}$ in the original Kruskal
gauge. In the semiclassical Eddington gauge the horizon is located
at the asymptotic boundary $\tilde{u} \to \infty$.

Recall that in the classical solution (with the same initial data)
the horizon is located at $U=0$. The inclusion of semiclassical effects
thus shifts the horizon in $U$ (or $\tilde{U}$) by an amount $qe^{-v_{0}}$.

\subsection{Behavior of $R$ and $S$ at the horizon}

The behavior of $R$ along the horizon is obtained by setting $H_{0}=H=0$
in Eq. (\ref{eq:Rfinal}), yielding $m_{v}+q[\ln(m_{v}/M_{0})+1]$.
However, the term $q\ln(m_{v}/M_{0})$ in this expression cannot be
trusted, as may be deduced from simple error estimate. To this end
we evaluate the error in $R_{,v}$ as a function of $v$ along the
horizon, using semiclassical Eddington coordinates for simplicity.
We do this by integrating the local error in $R_{\tilde{u}v}$ along
a line $v=const<0$, from $\tilde{u}=-\infty$ (PNI) to $\tilde{u}=\infty$
(horizon). From the discussion in Sec. \ref{sec:Verification} it
follows that along such a $v=const$ line the local error gets a maximal
value of order $q^{2}/R$$\sim q^{2}/m_{v}$ at intermediate $\tilde{u}\sim-v$
values, and it decays exponentially in $\tilde{u}$ in both directions.
The effective integration interval (in $\tilde{u}$) is of order unity,
hence the integrated error in $R_{,v}$ is also $\sim q^{2}/m_{v}=q/|v|$.
Consequently, the integrated error in $R(v)$ along the horizon is
of order $\sim q\ln(v/v_{0})=q\ln(m_{v}/M_{0})$. 
\footnote{It should be emphasized that despite this $O(q)$ integrated error,
it is crucial to keep the $O(q)$ term in the approximate expression
(\ref{eq:Rfinal}) for $R$. Without this term, the local error in
$R_{\tilde{u}v}$ will grow from $O(q^{2})$ to $O(q)$.} 
We therefore re-write the above result for $R(v)$ as 
\footnote{In fact one can use the constraint equation for $T_{vv}$ to obtain
the correct coefficient of the log term in Eq. (\ref{eq:Rhorizon}),
but this is beyond the scope of the present paper.}
\begin{equation}
R^{hor}(v)=m_{v}+q+O\left(q\ln\frac{m_{v}}{M_{0}}\right)=-qv+q+O\left(q\ln\frac{v}{v_{0}}\right).
\label{eq:Rhorizon}\end{equation}

Next we analyze the behavior of $S$ in the horizon's neighborhood,
in the semiclassical Eddington gauge, using Eq. (\ref{eq:Sfinal})
which now reads $S_{[\tilde{u}]}=\ln(-H_{,\tilde{u}})$. To this end
we divide Eq. (\ref{eq:HduDIF}) by $H_{,u}$, substitute $H=0$ in
the right-hand side, and re-write this equation as \begin{equation}
\frac{d}{dv}\ln\left(-H_{,\tilde{u}}\right)=1-\frac{1}{v}.\label{eq:SdvHOR}\end{equation}
The initial condition at $v=v_{0}$ is obtained from Eq. (\ref{eq:H0edditngton})
which yields $\ln(-H_{0,\tilde{u}})=v_{0}-\tilde{u}$. Integrating
Eq. (\ref{eq:SdvHOR}) we find \[
S_{[\tilde{u}]}=v-\tilde{u}+\ln\frac{v_{0}}{v}.\]
The error estimate for $S$ parallels the one carried out above for
$R$. It yields an accumulated error $\sim(q/m_{v})^{2}=v^{-2}$ in
the value of $S_{,v}$ at the horizon, and hence an integrated error
of order $\sim1/|v|=q/m_{v}$ in $S$ itself. 
\footnote{To be more precise, the integrated error is $\sim(1/|v|-1/|v_{0}|)=(q/m_{v}-q/M_{0})$.} 
We therefore re-write our result as \begin{equation}
S_{[\tilde{u}]}^{hor}=v-\tilde{u}+\ln\frac{v_{0}}{v}+O(1/v)=v-\tilde{u}+\ln\frac{M_{0}}{m_{v}}+O(q/m_{v}).\label{eq:Shorizon}\end{equation}

Another way to obtain this result is by integrating Eq. (\ref{eq:eql})
for $l(v)$. In the horizon's neighborhood this equation reduces to
$l_{,v}=1-1/v$, which is easily integrated (with the appropriate
initial conditions) to yield $l=v-\tilde{u}+\ln(v_{0}/v)$, or \begin{equation}
H^{hor}=\frac{v_{0}}{v}e^{v-\tilde{u}}.\label{eq:Hhor}\end{equation}
Differentiating this expression with respect to $\tilde{u}$ and substituting
in Eq. (\ref{eq:Sfinal}), one recovers Eq. (\ref{eq:Shorizon}).
\footnote{The alternative approximation $S^{alt}$ yields the same result. To
see this one substitutes Eq. (\ref{eq:Hhor}) in (\ref{eq:SaltEDext})
(and omit the unsecured $O(q)$ term).}

\subsection{The apparent horizon}

Motivated by the terminology used for conventional 4D spherically-symmetric
BHs, we define the apparent horizon to be the locus of the points
where $R_{,v}=0$ (or equivalently, $\phi_{,v}=0$). By virtue of
Eq. (\ref{eq:Rfinal}) this implies $(H+m_{v})_{,v}=0$, or $H_{,v}=q$.
Utilizing Eq. (\ref{eq:HdvSUM}), and restricting the analysis to
first order in $q$, we find that at the apparent horizon $H\cong q$
is satisfied.

We shall consider here the properties of the apparent horizon during the macroscopic phase $m_{v}\gg q$, namely $v \ll -1$. 
Throughout this phase $H^{hor} \ll R\cong m_{v}$ and hence the horizon approximation $H\cong H^{hor}$ applies. Setting $H\cong q$ in Eq. (\ref{eq:Hhor}) we find the apparent-horizon's location 
\begin{equation}
\tilde{u}\cong v+\ln(v_{0}/v)-\ln q.
\label{AH}
\end{equation}
Thus, $d \tilde{u}/dv=1-1/v \approx 1$ along the apparent horizon. We find that the apparent horizon is a timelike line, which is approximately vertical in the ($\tilde{u},v $) coordinates. In Fig. 1 the apparent horizon is denoted "AH".

Consider now the behavior of $H$ and $R$ along an outgoing null
geodesic located outside the BH though fairly close to the event horizon---namely,
sufficiently-large fixed $\tilde{u}$. To be more specific, let us
assume that $v_{0}<\tilde{u}<0$, such that $\tilde{u}\ll-1$ but
$\tilde{u}-v_{0}\gg 1$. (As a typical example one may take $\tilde{u}\approx v_{0}/2$;
Recall that throughout this paper we assume $v_{0}\ll-1$, 
corresponding to a macroscopic BH.) 
Then $H_{0}=e^{v_{0}-\tilde{u}}$ is exponentially small and
may be neglected in Eq. (\ref{eq:Rfinal}). Initially, near $v=v_{0}$,
$H$ is also exponentially small and therefore (neglecting terms of
order $\sim q$ compared to $m_{v}$), $R\approx m_{v}$. In this
range $R$ shrinks linearly in $v$, like $m_{v}$ itself (and like
$R^{hor}$). During this stage $H\cong H^{hor}$ grows (approximately)
exponentially in $v$, but initially this does not have much effect
on $R(v)$ because $H$ is still too small. However at some point
the exponentially-growing $H$ starts to slow the decrease rate of
$R(v)$. When $H$ approaches $q$ this exponential growth just balances
the linear decrease of $m_{v}$. This is the point of intersection
with the apparent horizon. Note that at this point $R$ is still $\approx m_{v}$,
the difference being $O(q)$. 
Soon afterwards the exponentially-growing $H$ overtakes $m_{v}$. 

On the other hand, for earlier outgoing geodesics with sufficiently
large $H_{0}$, the exponential growth of $H$ will dominate over $m_{v}$ everywhere,
and $R$ will grow monotonically all the way from $v_{0}$ to $\infty$.
Since $H_{,v}>0$ outside the BH, and the apparent horizon satisfies $H\cong q$,
this monotonic growth of $R$ will occur at the outgoing null geodesics
for which $H_{0}>q$. 

The evaporating-BH spacetime may thus be divided into three domains
in $u$. For concreteness let us use here the coordinate $\tilde{U}$
to characterize these domains: In the early domain, $\tilde{U}<\tilde{U}_{ah}$,
$R$ grows monotonically with $v$ throughout $v_{0}<v<\infty$. From
Eq. (\ref{eq:H0(shiftedU)}) we find (equating $H_{0}$ to $q$ as
explained above)\begin{equation}
\tilde{U}_{ah}=-qe^{-v_{0}}.\label{eq:Uah}\end{equation}
In the second domain, $\tilde{U}_{ah}<\tilde{U}<0$, $R$ first decreases
along an outgoing null ray (taking values fairly close to $m_{v}$)
until it intersects the apparent horizon at certain $v<0$, and then
$R$ starts to increase with $v$. In the third domain $\tilde{U}>0$
(the BH interior), $H$ is everywhere decreasing (and the same for
$m_{v}$), therefore $R$ decreases monotonically until the outgoing
null ray hits the spacelike singularity. 

From the discussion above it follows that at a given $v<0$ the event
and apparent horizons have roughly the same $R$ value (the
difference being approximately $q$). In other words, the apparent
and event horizons shrink (in $R$) in the same standard rate, $dR/dv=-q$. 

It may be interesting to compare three points along the worldline
of the collapsing shell: (1) The intersection of $v=v_{0}$ with the
apparent horizon ($\tilde{U}=\tilde{U}_{ah}$); (2) its intersection
with the event horizon ($\tilde{U}=0$); and (3) its intersection
with the (would-be) event horizon of the \emph{classical} CGHS BH
(namely $U=0$, or $\tilde{U}=qe^{-v_{0}}$). One finds that these
three points are equally-separated in $\tilde{U}$ (or in $U$), the
separation being $qe^{-v_{0}}$. We find it more illuminating, however,
to express these three points by their respective $R=R_{0}(u)$ values.
Noting that \[
R_{0}(\tilde{U})=M_{0}-\tilde{U}e^{v_{0}}+q,\]
one finds that $R_{0}$ is $M_{0}+2q$ at point 1, $M_{0}+q$ at point
2, and $M_{0}$ in point 3, so these points are equally-separated
in $R$ too. Notice, however, that for a macroscopic BH ($M_{0}\gg q$,
which we assume throughout) the relative separation $q/R_{1,2,3}\approx q/M_{0}$
is $\ll1$.

\subsection{Influx at the horizon}

We proceed now to calculate $T_{vv}$ at the horizon, using the constraint
equation (\ref{eq:Rvv}) which now reads\[
T_{vv}^{hor}=R_{,v}^{hor}S_{,v}^{hor}-R_{,vv}^{hor}.\]
We shall restrict here the calculation to the leading order, namely
first order in $q/m_{v}$. From Eqs. (\ref{eq:Rhorizon},\ref{eq:Shorizon})
we find $S_{,v}^{hor}=1+O(q/m_{v})$, $R_{,v}^{hor}=-q+O(q^{2}/m_{v})$,
and $R_{,vv}^{hor}=O(q^{3}/m_{v}^{2})$ will not contribute. We obtain
at the leading order\begin{equation}
T_{,vv}^{hor}=-q=-\frac{K}{4}=-\frac{N}{48}.\label{eq:TvvHOR}\end{equation}
Thus, in the macroscopic limit the influx into the horizon is constant
and independent of the mass---a well-known result for a two-dimensional
BH \cite{CGHS}. 

In principle it is possible to use Eq. (\ref{eq:Tvv}) for $\hat{T}_{vv}$
to obtain the first-order correction to the fixed influx (\ref{eq:TvvHOR})
(and thereby to fix the $O[q\ln(m_{v}/M_{0})]$ term in $R^{hor}$),
but this is beyond the scope of the present paper.

\section{Null infinity \label{sec:PFNI}}

\subsection{Past null infinity \label{sub:PNI}}

The limit $U \to -\infty$ (also $\tilde{U},\tilde{u} \to -\infty$)
and finite $v$ corresponds to PNI. In this asymptotic
boundary $m_{v}$ is finite but $H$ diverges (this immediately follows
from the divergence of $H_{0}\propto-\tilde{U}$, combined with the
monotonic growth of $|H|$ with $v$). At this limit the ODE (\ref{eq:HdvSUM})
for $H$ reduces to $H_{,v}\cong H+q$ and its solution, corresponding
to the initial conditions (\ref{eq:H0Usum}), is \begin{equation}
H^{pni}\cong-Ue^{v}-q.\label{eq:Hpni}\end{equation}
The expression (\ref{eq:Sfinal}) for $S$ then yields\begin{equation}
S_{[U]}^{pni}\cong v.\label{eq:Spni}\end{equation}
(For the other gauges we find $S_{[\tilde{U}]}^{pni}=v$ and $S_{[\tilde{u}]}^{pni}=v-\tilde{u}$.)
For $R$ Eq. (\ref{eq:Rfinal}) yields $R\cong$$m_{v}-Ue^{v}+q\ln(H/H_{0})$,
and setting $\ln(H/H_{0})\cong v-v_{0}$ we obtain \begin{equation}
R^{pni}\cong M-Ue^{v}.\label{eq:Rpni}\end{equation}
These expressions for $R^{pni}$ and $S^{pni}$ properly match the
desired initial conditions (\ref{eq:PNIdata}) at PNI. 

By assumption the influx $T_{vv}$ should vanish at PNI.
We would like to verify this by applying the constraint equation $T_{vv}=R_{,v}S_{,v}-R_{,vv}$
to our approximate solution and taking the limit $U \to -\infty$.
Doing so we observe that Eqs. (\ref{eq:Spni},\ref{eq:Rpni}) yield
$R_{,v}^{pni}=R_{,vv}^{pni}=-Ue^{v}$ and $S_{,v}^{pni}=1$, which
yields the desired result\begin{equation}
T_{vv}^{pni}=0.\label{eq:TvvPNI}\end{equation}

Note the following subtlety, however: Because $R_{,v}^{pni}\propto U$
diverges at PNI, the term $R_{,v}^{pni}S_{,v}^{pni}$ might receive
a nonvanishing contribution from $O(1/U)$ corrections to $S$, if
such existed. To address this issue we must carry the calculation
of $S_{[U]}^{pni}$ to order $1/U$. This in turn requires a more
detailed examination of $H^{pni}$. At PNI the ODE
(\ref{eq:HdvSUM}) takes the form $H_{,v}=H+q+O(1/H)$, and the last
term introduces $O(1/U)$ corrections to $H$, namely $H^{pni}\cong-Ue^{v}-q+O(1/U).$
However, this only leads to $O(1/U^{2})$ corrections in $S_{[U]}^{pni}$,
which leave Eq. (\ref{eq:TvvPNI}) intact.

\subsection{Future null infinity \label{sub:FNI}}

The limit $v \to \infty$ (for $\tilde{U}<0$) corresponds to
FNI. The analysis of this asymptotic region is more
complicated than that of PNI, because this time we have to integrate
the ODE (\ref{eq:HdvSUM}) through strong-field regions (where $H$
is comparable to $m_{v}$). The thorough investigation of this asymptotic
region is beyond the scope of the present paper, and we hope to address
it elsewhere. Here we shall merely mention two key properties: (1)
Spacetime is asymptotically-flat; In other words, for appropriate
choice of $u$-coordinate, which we denote $\hat{u}$ (and which does
not exactly coincide with $\tilde{u}$), $\rho_{[\hat{u}]}$ vanishes
at $v \to \infty$. (2) At the leading order one obtains a constant,
mass-independent, Hawking outflux $T_{\hat{u}\hat{u}}=q=N/48$ (a
well-known result \cite{CGHS}).

\section{Summary \label{sec:Summary}}

Our approximate solution for the semiclassical variables $R$ and
$S$ is described in Eqs. (\ref{eq:Rfinal},\ref{eq:Sfinal}). The
determination of the original CGHS variables $\phi,\rho$ from $R$
and $S$ is straightforward. The function $H$ involved in this solution
is determined by the ODE (\ref{eq:HdvSUM}), along with the initial
conditions (\ref{eq:H0SUM}) or (\ref{eq:H0Usum}). Approximate expressions
for $H$ are given in Eqs. (\ref{eq:h1},\ref{eq:h1ed}). 

In the CGHS formalism all semiclassical effects originate from a term
in the action which is proportional to $K\equiv N/12$. Our approximation
scheme is restricted to macroscopic BHs, namely, those with original
mass $M_{0}\gg q\equiv K/4$. Furthermore, it only applies to spacetime
regions where $R\gg q$. It nevertheless holds both outside and inside the BH,
though obviously not too close to the singularity (where the condition
$R\gg q$ is violated). 

Throughout this paper we use double-null coordinates ($u,v$). Our
construction explicitly preserves the gauge-freedom in $u$, though not
in $v$. Our coordinate $v$ coincides with the affine parameter along
PNI. (We have set the origin of $v$ such that the collapsing shell
is placed at $v=v_{0}\equiv-M_{0}/q$.) For the $u$-gauge we find
two particularly useful choices: the shifted-Kruskal coordinate $\tilde{U}$,
and the semiclassical Eddington coordinate $\tilde{u}$ (both defined
in Sec. \ref{sec:Gauges}).

At the horizon's neighborhood we find that $R$ shrinks linearly with
$v$, as may be anticipated: $R\cong-qv+q$. The other variable $S$
behaves as $S\cong v+\ln(v_{0}/v)+c$, where $c$ is $v$-independent
though it depends upon the $u$-gauge being used. (For example, $c=0$
in the shifted-Kruskal gauge and $c=-\tilde{u}$ in the semiclassical
Eddington gauge.) The term $v$ merely reflects the classical behavior
of $S$ at the horizon, whereas the log term is of semiclassical origin. 

Our approximate solution is "locally first-order accurate";
Namely, the deviation of $R_{,\tilde{u}v}$ and $S_{,\tilde{u}v}$
from their respective values, as specified in the evolution equations
(\ref{eq:Requation},\ref{eq:Sequation}) or 
(\ref{eq:RuvSTANDARD},\ref{eq:SuvSTANDARD}), is proportional to
$q^{2}$. [More specifically, the deviations in both $R_{,\tilde{u}v}/R$
and $S_{,\tilde{u}v}$ are $\sim(q/R)^{2}$ at typical (off-horizon)
strong-field regions, and smaller elsewhere.] On the other hand,
the accumulated error in $S$, and accumulated relative error in $R$,
are of first order in $q/R$. This change in the power of $q$ is
because the effective accumulation intervals (in $v$ and/or $\tilde{u}$)
are typically of order of the evaporation time, which is $M_{0}/q$. 

It will be interesting to check this approximate solution against
numerical simulations of the CGHS field equations. Such numerical
simulations are currently being conducted by several groups \cite{Dori,Pretorius}.
A preliminary comparison with the numerical results \cite{Dori}
shows nice agreement, though a more comprehensive check still needs
to be carried out.

\section*{ACKNOWLEDGEMENTS}

I would like to thank Liora Dori, Fethi Ramazanoglu and Frans Pretorius
for helpful discussions, and for sharing their numerical results.
I am especially grateful to Abhay Ashtekar for his warm hospitality
and for many interesting and fruitful conversations.

\appendix

\section{Approximate solution for $H(v;u)$ \label{sec:Appendix-H}}

A useful approximate solution of the ODE (\ref{eq:HdvSUM}) for $H$
is \begin{equation}
H(v;\tilde{U})\cong-\tilde{U}e^{v}\frac{\ln(e^{-v_{0}}-\tilde{U}/q)}{\ln(e^{-v}-\tilde{U}/q)},\label{eq:h1}\end{equation}
where $\tilde{U}\equiv U+qe^{-v_{0}}$ is the shifted Kruskal coordinate
defined in Sec. \ref{sec:Gauges}. This expression holds both inside
and outside the BH (corresponding to $\tilde{U}>0$ and $\tilde{U}<0$
respectively).

Note that the initial condition $H_{0}(\tilde{U})=-\tilde{U}e^{v_{0}}$ at $v=v_{0}$ is precisely satisfied.
Also recall that the dependence of $H$ on $\tilde{U}$ only emerges through
this initial condition at the shell. 
It is thus possible to substitute $\tilde{U}=-H_{0}e^{-v_{0}}$ in
Eq. (\ref{eq:h1}), to obtain an expression for $H$ in terms of $H_{0}(u)$
but with no explicit reference to any specific $u$-gauge. 

We also translate the above expression for $H$ to the semiclassical
Eddington coordinate $\tilde{u}$, valid in the external world ($\tilde{U}<0$):
\begin{equation}
H(v;\tilde{u})\cong e^{v-\tilde{u}}\frac{\ln(e^{-v_{0}}+e^{-\tilde{u}}/q)}{\ln(e^{-v}+e^{-\tilde{u}}/q)}.
\label{eq:h1ed}\end{equation}

Preliminary error estimate suggests the following behavior of the
relative error in the above expression for $H$: We assume $v_{0}\ll-1$
throughout. For $u\sim v_{0}$ or $v_{0}<u\ll-1$ the error scales
as $1/u$, though with some logarithmic corrections. This applies
to both (i) the relative error in $H$ itself, and (ii) the relative
local error in $H_{,v}$ [compared to its respective value in the
ODE (\ref{eq:HdvSUM})]. For $u<v_{0}$ the error further scales
as $e^{u-v_{0}}$ and quickly becomes negligible.

\section{Calculating the constant $p$ \label{sec:Appendix-Const}}

In this Appendix we determine the constant $p$ in Eq. (\ref{eq:H0gen})
by analyzing the asymptotic behavior of $R$ at PNI and comparing it to the desired initial conditions.

PNI is characterized by finite $v$ but $U \to -\infty$,
yielding finite $m_{v}$ but $H \to \infty$ (the latter follows
from the divergence of $H_{0}$ combined with the monotonic growth
of $|H|$ with $v$). Therefore, in the ODE (\ref{eq:HdvSUM}) we
may replace the term $H-qv$ in the denominator by $H$. We are left
with the ODE $H_{,v}\cong H+q$, whose general solution is\[
H(u,v)=-q+[H_{0}(u)+q]e^{v-v_{0}}.\]
Using the general expression (\ref{eq:H0gen}) for $H_{0}(U)$ we
get \begin{equation}
H(U,v)=[-Ue^{v_{0}}+(p+1)q]e^{v-v_{0}}-q.\label{eq:HpniGEN}\end{equation}
We now substitute this in Eq. (\ref{eq:W1}) for $W$ 
(or more precisely, $W^{in}$; see below). 
In doing so, we may approximate $q\ln(H+m_{v})$ at PNI by $q\ln H$, which by virtue of Eq. (\ref{eq:HpniGEN}) may be approximated by $q[v+\ln(-U)]=-m_{v}+q\ln(-U)$,
omitting $O(q^{2})$ contributions. (The same will apply to terms
like $q\ln W$ and $q\ln R_{0}$ below.) We obtain \begin{equation}
W^{in}(U,v)=[-Ue^{v_{0}}+(p+1)q]e^{v-v_{0}}+2m_{v}-2q-q\ln(-U).\label{eq:Winp}\end{equation}
We used here the symbol $W^{in}$ to make it clear that this is the
$W$-function \emph{associated with the ingoing Vaidya-like core solution}
(as opposed to the full-$W$ function, considered below; See discussion in Sec. \ref{sec:perturb}). At $v=v_{0}$
this yields \[
W^{in}(U,v_{0})=-Ue^{v_{0}}+2M_{0}-q\ln(-U)+(p-1)q.\]
This should be compared to the full-$W$ initial data at $v=v_{0}$,
given by Eqs. (\ref{eq:W0U},\ref{eq:R0U}):\[
W_{0}(U)=M_{0}-Ue^{v_{0}}-2q[v_{0}+\ln(-U)]=3M_{0}-Ue^{v_{0}}-2q\ln(-U).\]
[Here, again, in processing the term $2q\ln R_{0}=2q\ln(M_{0}-Ue^{v_{0}})$
in Eq. (\ref{eq:W0U}) we have neglected $M_{0}$ compared to $Ue^{v_{0}}.$]
The difference is \begin{equation}
\delta W(U)\equiv W_{0}(U)-W^{in}(U,v_{0})=M_{0}-q\ln(-U)-(p-1)q.\label{eq:deltaW}\end{equation}

The full function $W(U,v)$ is obtained by simply adding the outgoing
perturbation $\delta W(U)$ to the ingoing solution $W^{in}(U,v)$
(this is shown in detail in Sec. \ref{sec:perturb}, based on linear
perturbation analysis). Summing Eqs. (\ref{eq:Winp},\ref{eq:deltaW})
we obtain for the full-$W$ function near PNI \[
W=[-Ue^{v_{0}}+(p+1)q]e^{v-v_{0}}+M_{0}+2m_{v}-2q\ln(-U)-(p+1)q.\]
Finally we transform from $W$ to $R$, using Eq. (\ref{eq:WtoR}).
The term $2q\ln W$ therein [in which we approximate $\ln W\cong v+\ln(-U)$
as explained above] just cancels the terms $2m_{v}-2q\ln(-U)$ in
the last expression for $W$, and we obtain\[
R=M_{0}-Ue^{v}+(p+1)q(e^{v-v_{0}}-1).\]
This expression should agree with the presumed PNI initial conditions
(\ref{eq:PNIdata}), namely $R=M_{0}-Ue^{v}$, which dictates $p=-1$.


\begin{thebibliography}{33}
\bibitem{notation} See Refs. \cite{gorbonos,Ori-flux}; To connect to the general formalism of flux-conserving systems presented therein one should set the notation $W \to \bar{R}$, $Z \to \bar{S}$ for Ref. \cite{Ori-flux}, and without the over-bars for Ref. \cite{gorbonos}.
\end{thebibliography}
\end{document}